\documentclass[prd,nofootinbib]{revtex4}
\usepackage{amssymb}
\usepackage{graphicx}
\usepackage{latexsym}
\usepackage{color}

\begin{document}

\title{Constraints on Primordial Non-Gaussianity from Large Scale Structure Probes}

\author{Jun-Qing Xia${}^1$}
\email{xia@sissa.it}
\author{Carlo Baccigalupi${}^{1,2,3}$}
\email{bacci@sissa.it}
\author{Sabino Matarrese${}^{4,5}$}
\email{sabino.matarrese@pd.infn.it}
\author{Licia Verde${}^6$}
\email{liciaverde@icc.ub.edu}
\author{Matteo Viel${}^{2,3}$}
\email{viel@oats.inaf.it}

\affiliation{${}^1$Scuola Internazionale Superiore di Studi
Avanzati, Via Bonomea 265, I-34136 Trieste, Italy}

\affiliation{${}^2$INAF-Osservatorio Astronomico di Trieste, Via
G.B. Tiepolo 11, I-34131 Trieste, Italy}

\affiliation{${}^3$INFN/National Institute for Nuclear Physics, Via
Valerio 2, I-34127 Trieste, Italy}

\affiliation{${}^4$Dipartimento di Fisica ``G. Galilei",
Universit\`a di Padova, Via Marzolo 8, I-35131 Padova, Italy}

\affiliation{${}^5$INFN, Sezione di Padova, Via Marzolo 8, I-35131
Padova, Italy}

\affiliation{${}^6$ICREA (Instituci\'o Catalana de Recerca i Estudis
Avan\c{c}ats) and Instituto de Ciencias del Cosmos,(ICC-UB-IEEC)
Universidad de Barcelona, Marti i Franques 1, 08028, Barcelona,
Spain}

\date{\today}

\begin{abstract}

In this paper we measure the angular power spectra $C_\ell$ of three
high-redshift large-scale structure probes: the radio sources from
the NRAO VLA Sky Survey (NVSS), the quasar catalogue of Sloan
Digital Sky Survey Release Six (SDSS DR6 QSOs) and the MegaZ-LRG
(DR7), the final SDSS II Luminous Red Galaxy (LRG) photometric
redshift survey.  We perform a global analysis of the constraints on
the amplitude of primordial non-Gaussianity from these angular power
spectra, as well as from their cross-correlation power spectra with
the cosmic microwave background (CMB) temperature map. In
particular, we include non-Gaussianity of the type arising from
single-field slow roll, multifields, curvaton (local type), and
those which effects on the halo clustering can be described by the
equilateral template (related to higher-order derivative type
non-Gaussianity) and by the enfolded template (related to modified
initial state or higher-derivative interactions).  When combining
all data sets, we obtain limits of $f_{\rm NL}=48\pm20$, $f_{\rm
NL}=50\pm265$ and $f_{\rm NL}=183\pm95$ at $68\%$ confidence level
for local, equilateral and enfolded templates, respectively.
Furthermore, we explore the constraint on the cubic correction
$g_{\rm NL}\phi^3$ on the bias of dark matter haloes and obtain a
limit of $-1.2\times10^5<g_{\rm NL}<11.3\times10^5$ at 95\%
confidence level.

\end{abstract}

%\pacs{}

\maketitle

\section{Introduction}\label{int}

The physical mechanisms responsible for the generation of primordial
perturbations seeding present-day large-scale structure, may leave
their imprint in the form of small deviations from a Gaussian
distribution of the primordial perturbations. Searches for
primordial non-Gaussianity can thereby provide key information on
the origin and evolution of cosmological structures (e.g., ref.
\cite{Komatsuwhitepaper} and references therein). Although the
standard single-field, slow-roll, canonical kinetic energy and
adiabatic vacuum state inflation generates very small
non-Gaussianity, any inflationary model that deviates from this may
entail a larger level of it (refs.~\cite{bmr04,Komatsu2010} and
references therein).

Deviations from Gaussian initial conditions are often taken to be of
the so-called local type and are parameterized by the constant
dimensionless parameter {$f_{\rm NL}$} \cite{Salopekbond90,
Ganguietal94, VWHK00, KS01, BabichCreminelliZaldarriaga}:
\begin{equation}
\Phi=\phi+f_{\rm NL}\left(\phi^2-\langle\phi^2\rangle\right)~,
\label{eq:fnl}
\end{equation}
where $\Phi$ denotes Bardeen's gauge-invariant potential (evaluated
deep in the matter era) and $\phi$ is a Gaussian random field
\footnote{In the literature there are two conventions: in the large
scale structure (LSS) convention $\Phi$ is linearly extrapolated to
$z=0$, while in the cosmic microwave background (CMB) convention it
is evaluated deep in the matter era.  Thus, $f^{\rm LSS}_{\rm NL} =
[g(z=\infty)/g(z=0)] f^{\rm CMB}_{\rm NL}\sim1.3f^{\rm CMB}_{\rm
NL}$, where $g(z)$ denotes the $\Lambda$-induced linear growth
suppression factor. In this paper we will use the CMB convention.}.

Recently, a method for constraining non-Gaussianity from large-scale
structure surveys has been proposed \cite{DDHS08,MV08}, which
exploits the fact that the clustering of dark matter halos --where
galaxies form-- on large scales is modified in a scale-dependent way
by the presence of even small amount of non-Gaussianity. In
particular, a non-Gaussianity described by eq.~(\ref{eq:fnl}),
introduces a scale-dependent boost (for $f_{\rm NL}>0$ and a
suppression for $f_{\rm NL}<0$) of the halo power spectrum
proportional to $1/k^2$ on large scales ($k<0.03\,h/$Mpc), which
evolves roughly as $(1+z)$.  Large-Scale Structure (LSS) surveys
covering large volumes are needed to access the scales where the
signal arises (e.g., ref. \cite{melita} and references therein). The
possibility of using high redshift data to constrain non-Gaussianity
has also been addressed by means of hydrodynamic simulations of the
Intergalactic Medium \cite{viel09}.  Among the many
currently-available tracers of the LSS, the radio sources from the
NRAO VLA Sky Survey (NVSS) \cite{Condon:1998iy} and the quasar
catalogue of the Sloan Digital Sky Survey Release Six (SDSS DR6
QSOs) \cite{richards09} are particularly interesting since they span
large volumes extending out to substantial redshifts
\cite{hoetal08}. Indeed these source samples were shown to provide
tight constraints on primordial non-Gaussianity by the
cross-correlation measurements between CMB temperature fluctuations
and the LSS number density which could be used to detect the late
Integrated Sachs-Wolfe (ISW) \cite{ISW} effect
\cite{Slosar,afshorditolley,xiaACF,xiaCCF}.

In recent papers \cite{xiaACF,xiaCCF} it was shown that the observed
NVSS and SDSS DR6 QSOs Auto-Correlation Function (ACF) and
Cross-Correlation Function (CCF) hint at a positive value of $f_{\rm
NL}$ at more than $2\,\sigma$. This is because the ACF is found to
be still positive on angular scales $\theta > 4^\circ$, which, for
the median source redshift ($z_{\rm m}\simeq 1$), correspond to
linear scales where the correlation function should be negative if
the density fluctuation field is Gaussian. A positive {$f_{\rm NL}$}
adds power on large angular scales, accounting for the observed ACF.

In this paper, we will revisit the constraints on the primordial
non-Gaussianity from these two LSS tracers using the power spectrum
$C_\ell$, instead of the correlation function $w(\theta)$.  While it
is true that the angular correlation function and the angular power
spectrum are spherical harmonic pairs and thus in principle they
should enclose the same physical information, each of the two
measurements is affected by systematic errors in different ways. For
example the correlation function is less affected by the survey mask
than the power spectrum but error estimation for the power spectrum
is much easier then for the correlation function. A direct
comparison of parameter constraints obtained via the correlation
function and the power spectrum analyses offer therefore a way to
quantify possible systematic effects. We will also use the recently
updated SDSS II LRG photometric redshift survey, MegaZ-LRG (DR7)
\cite{LRG7} to constrain the primordial non-Gaussianity. Ref.
\cite{Slosar} used the LRG angular power spectrum for a sample close
to the SDSS DR3 to constrain $f_{\rm NL}$. Here we revisit the
analysis using an improved bias model, improved LRG catalog, updated
knowledge of the sources redshift distribution, up-to-date CMB maps
and complementary analysis methods.

Furthermore, we explore the constraints on physically motivated
primordial non-Gaussianity shapes that are different from the local
case. We consider non-Gaussianities which effects on the halo
clustering is well described by the equilateral and enfolded
templates.  We consider the angular power spectra of these three LSS
tracers, as well as their cross-correlation power spectra with the
CMB temperature fluctuation. Finally, we consider the cubic
correction on the halo bias, which can be motivated in scenarios
like the curvaton model, in which a large cubic correction can be
produced while simultaneously keeping the $f_{\rm NL}$ correction
small \cite{Gnlpaper,loverdeetal08}.

The structure of the paper is as follows. In section \ref{theory} we
review the effects of primordial non-Gaussianity on the power
spectrum $C_\ell$. Section \ref{data} contains the analysis of power
spectrum $C_\ell$ for NVSS radio sources, SDSS DR6 QSOs and
MegaZ-LRG. In section \ref{method} we present the method used to
derive constraints on $f_{\rm NL}$. Section~\ref{result} contains
our main results. We conclude with a discussion in section
\ref{summary}.

\section{Effects of Primordial Non-Gaussianity}\label{theory}

The effects of non-Gaussianity on the source clustering properties
arise because a non-zero $f_{\rm NL}$ affects the halo mass function
and enhances the halo clustering on large scales. The second effect
is the dominant one on large scales.

In the presence of primordial non-Gaussianity, the mass function
$n_{\rm NG}(M,z,f_{\rm NL})$ can be written in terms of the Gaussian
one $n_{\rm G}^{\rm sim}(M,z)$, for which a good fit to the results
of simulations is provided by e.g., the Sheth-Tormen formula
\cite{shethtormen}, multiplied by a non-Gaussian correction factor
\cite{mvj,VJKM01,loverdeetal08} \footnote{Although attempts have
been made to derive directly an expression for the non-Gaussian mass
function \cite{MR10,D'Amico:2010ta}.}:
\begin{eqnarray}
R_{\rm NG}(M,z,f_{\rm NL})=1+\frac{\sigma^2_{\rm M}}{6\delta_{\rm
ec}(z)}\left[\!S_{\rm 3,M}\!\left(\!\frac{\delta^4_{\rm
ec}(z)}{\sigma^4_{\rm M}} - 2\frac{\delta^2_{\rm
ec}(z)}{\sigma^2_{\rm M}} - 1\!\!\right)\! +\!\frac{dS_{\rm
3,M}}{d\!\ln\!{\sigma_{\rm M}}}\!\left(\!\frac{\delta^2_{\rm
ec}(z)}{\sigma^2_{\rm M}}- 1\!\!\right)\!\right]~,
\end{eqnarray}
where the normalized skewness of the density field, $S_{\rm 3,M}$,
is $\propto f_{\rm NL}$, $\sigma_{\rm M}$ denotes the rms of the
dark matter density field linearly extrapolated to $z=0$ and
smoothed on scale $R$ corresponding to the Lagrangian radius of a
halo of mass $M$, and $\delta_{\rm ec}$ is the critical density for
ellipsoidal collapse, calibrated on N-body simulations
\cite{grossi09,wagner1,desjacque09}. For high peaks ($\delta_{\rm
ec}/\sigma_{\rm M}\gg 1$) and small $f_{\rm NL}$, $\delta_{\rm ec}$
is slightly smaller than the critical density for spherical
collapse, $\delta_{\rm c}(z) = \Delta_{\rm c}(z)D(0)/D(z)$ where
$D(z)$ is the linear growth factor, and $\Delta_{\rm c}(z)\sim 1.68$
evolves very weakly with redshift.

The large-scale halo bias is also modified by the presence of
non-Gaussianity
\citep{DDHS08,MV08,grossi09,Wagner2,Smith:2009jr,gp10}:
\begin{equation}
b_{\rm NG}(z)-b_{\rm G}(z)\simeq2(b_{\rm G}(z)-1)f_{\rm
NL}\delta_{\rm ec}(z)\alpha_{\rm M}(k)~, \label{eq:nghalobias}
\end{equation}
where the factor $\alpha_{\rm M}(k)$ encloses the scale and halo
mass dependence. Here, we consider three types of non-Gaussianity.

In practice, we find that, on large scales, $\alpha_{\rm
M}(k)\propto 1/k^2$, is independent of the halo mass for the local
type. The factor $\alpha_{\rm M}(k)$ for the enfolded template is
proportional to $1/k$, while the equilateral template gives an
almost scale-independent $\alpha_{\rm M}(k)$. It is important to
note here that the equilateral and enfolded templates were created
to have factorizable expressions that gave a good description of the
actual bispectra, on average, over all configurations.  For the the
halo bias effect, nearly squeezed configurations dominate. As
discussed e.g., in Ref. \cite{Wagner2} the templates may not
necessarily offer a good approximation of the actual bispectra
generated by higher-order derivatives or vacuum-state modifications.
For all inflationary single-field models the bispectrum scales
either as $1/k$ or as $1/k^3$ in the squeezed limit (i.e. either
like the equilateral template or like the local case)
\cite{Wagner2}.  The orthogonal template instead scales like $1/k^2$
in the squeezed limit. Nevertheless it is interesting to consider
such an intermediate case: inflationary models that go beyond single
field can show a range of scaling in the squeezed limit (e.g., ref.
\cite{Chen10}).

We assume that the large-scale linear halo bias for the Gaussian
case is \cite{shethtormen}:
\begin{eqnarray}\label{eq:bST}
b_{\rm G}&=&1+\frac{1}{D(z_{\rm o})}\left[\frac{q\delta_{\rm
c}(z_{\rm f})}{\sigma^2_{\rm M}}-\frac{1}{\delta_{\rm c}(z_{\rm
f})}\right]+\frac{2p}{\delta_{\rm c}(z_{\rm f})D(z_{\rm
o})}\left\{1+\left[\frac{q\delta^2_{\rm c}(z_{\rm f})}{\sigma^2_{\rm
M}}\right]^p\right\}^{-1}~,
\end{eqnarray}
where $z_{\rm f}$ is the halo formation redshift, and $z_{\rm o}$ is
the halo observation redshift. As we are interested in massive
haloes, we expect that $z_{\rm f}\simeq z_{\rm o}$. Here, $q=0.75$
and $p=0.3$ account for the non-spherical collapse and are a fit to
numerical simulations (see also refs.
\cite{Mo96,Mo97,Scoccimarro01}). Finally, the weighted effective
halo bias is given by
\begin{equation}
b_{\rm NG}^{\rm eff}(M_{\rm min},z,k,f_{\rm
NL})=\frac{\int^\infty_{M_{\rm min}}b_{\rm NG}n_{\rm
NG}dM}{\int^\infty_{M_{\rm min}}n_{\rm NG}dM},
\end{equation}
$M_{\rm min}$ being the minimum halo mass hosting a source of the
kind we are considering.

\begin{figure}[htpb]
\begin{center}
\includegraphics[scale=0.4]{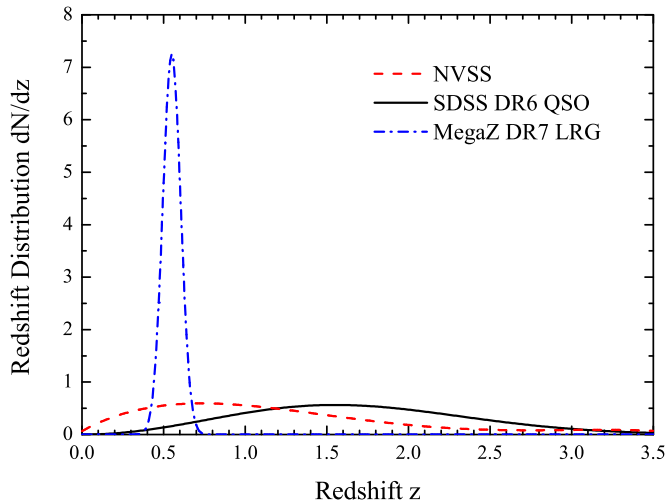}
\caption{Normalized redshift distributions, $dN/dz$, of the
different types of objects considered for our analysis. NVSS radio
sources (red dashed line), SDSS DR6 QSOs (black solid line), and
SDSS DR6 LRGs (blue dash-dotted line).} \label{fig:redshift}
\end{center}
\end{figure}

\section{Observed Power Spectrum}\label{data}

\subsection{Discrete Sources Maps}

In this subsection we describe three LSS probes, NVSS radio sources,
high redshift SDSS DR6 QSOs and MegaZ DR7 LRGs sample. In figure
\ref{fig:redshift} we show their redshift distribution, $dN/dz(z)$.
All distributions are normalized to unit integral.

\subsubsection{NVSS Radio Sources}\label{ACFdata}

The NVSS \cite{Condon:1998iy} offers the most extensive sky coverage
($82\%$ of the sky to a completeness limit of about 3 mJy at 1.4
GHz). We start by confining our analysis to NVSS sources brighter
than 10 mJy, since the surface density distribution of fainter
sources suffers from declination-dependent fluctuations
\cite{BlakeWall}. Density gradients in the NVSS catalog become
increasingly unimportant as the source brightness threshold is
increased. Figure 6 of ref. \cite{BlakeWall}, figure 1 of ref.
\cite{BlakeWall2} and figure 3 of ref. \cite{CHM10} showed
variations of NVSS source surface density as a function of
declination for sources with different flux thresholds. Dim sources
are strongly affected by density gradients, however this does not
appreciably happen for the brighter sources. Furthermore, in our
previous work we used the NVSS sources with different flux
thresholds $S \geq 10$ mJy and $S\geq20$ mJy to compute the ACF of
NVSS source and its CCF with CMB temperature fluctuations and found
that the obtained ACF and CCF results are stable \cite{xiaCCF}. Also
we mask the stripe $|b|<5^\circ$, where the catalog may be
substantially affected by Galactic emissions. Here, we have
explicitly checked that the choices of masked stripes with
$|b|<10^\circ$ and $|b|<20^\circ$ have negligible impact on our
final results. In order not to excessively reduce the sample, in our
analysis we consider the NVSS sources brighter than 10 mJy and the
masked strip with $|b|<5^\circ$ to reduce the contamination from
Galactic emissions. The NVSS source surface density at this
threshold is $16.9\,{\rm deg}^{-2}$ and the redshift distribution at
this flux limit has been recently determined by ref. \cite{Brookes}.
Their sample, complete to a flux density of 7.2 mJy, comprises 110
sources with $S_{1.4\rm GHz}\ge 10\,$mJy, of which 78 (71\%) have
spectroscopic redshifts, 23 have redshift estimates via the $K$--$z$
relation for radio sources, and 9 were not detected in the $K$ band
and therefore have a lower limit to $z$. We adopt here the smooth
description of this redshift distribution given by ref.
\cite{DeZotti10}, which is shown in the red dashed line of figure
\ref{fig:redshift}:
\begin{equation}\label{eq:zdis}
dN/dz=1.29+32.37z-32.89z^2+11.13z^3-1.25z^4~.
\end{equation}
The mean redshift of this sample is $\bar{z}=1.23$ and the redshift
range is $0<z<3.5$.

\subsubsection{SDSS DR6 Quasars}

The SDSS DR6 QSO catalog released by Ref. \cite{richards09} contains
about $10^6$ objects with photometric redshifts ranging from $0.065$
to $6.075$ over a total area of $8417\,{\rm deg}^2$ ($\sim20\%$ of
the whole sky). We refer the reader to ref. \cite{richards09} for a
detailed description of the object selection with the non-parametric
Bayesian classification kernel density estimator (NBC-KDE)
algorithm. We use the electronically-published table that contains
only objects with the ``good'' flag with values in the range
$[0,6]$. The higher the value, the more probable for the object to
be a real QSO (see \S\,4.2 of ref. \cite{richards09} for details).
Furthermore we restrict ourselves to the ``uvxts=1'', i.e. to QSOs
clearly showing a UV excess which should be a signature of a QSO
spectrum (in this case we have $N_{\rm qso}\approx6\times10^5$
QSOs). As we know, several systematics effects, including the
galactic extinction by dust, sky brightness, number of point sources
and poor seeing, could potentially affect both the observed
auto-/cross-correlation power spectra. In our previous work we
carefully checked for their contribution in calculations and in
particular we considered extinction and point sources contamination.
We found that the dominant systematic effect is the extinction
\cite{xiaISW,gianna06}. Therefore, in order to minimize the effect
of Galactic extinction on the observed QSO distribution, we use the
extinction mask $A_{\rm g} < 0.18$ only in our following analysis
\cite{xiaISW,myersetal06,gianna06}. These masks will remove about
$\sim$20\% of the considered area.

We fit the redshift distribution $dN/dz$ of the DR6 QSO sample with
a function of the form \cite{xiaISW} (black solid line of figure
\ref{fig:redshift}):
\begin{equation}
\frac{dN}{dz}(z)=\frac{\beta}{\Gamma(\frac{m+1}{\beta})}\frac{z^m}{z^{m+1}_0}
\exp\left[-\left(\frac{z}{z_0}\right)^\beta\right]~.\label{reddis}
\end{equation}
The best-fit values of the parameters are $m=2.00$, $\beta=2.20$,
$z_0=1.62$; the mean redshift of the sample is $\bar{z}\sim1.49$.

\subsubsection{MegaZ DR7 LRG}

We use the updated MegaZ LRG DR7 sample \cite{LRG7}, which contains
$\sim1.5\times10^6$ galaxies from the SDSS DR7 in the redshift range
$0.4 < z < 0.7$ with limiting magnitude $i < 20$. To reduce stellar
contamination, there is a variable of the MegaZ neural network
estimator $\delta_{\rm sg}$, defined such that $\delta_{\rm sg} = 1$
if the object is a galaxy, and $\delta_{\rm sg} = 0$ if it is a
star. For a conservative analysis, we choose a cut $\delta_{\rm sg}
> 0.2$, which is reported to reduce stellar contamination below 2\%
while keeping 99.9\% of the galaxies. In addition to the SDSS DR7
geometry mask, we also add two foreground masks to account for
seeing (removing pixels with median seeing in the red band larger
than 1.4 arcsec) and reddening (removing pixels with median
extinction in the red band $A_{\rm r} > 0.18$) \cite{gianna08}. The
redshift distribution function (blue dash-dotted line of figure
\ref{fig:redshift}) in this case is found directly from the
photometric redshifts that are given in the catalogue
\cite{LRGredshift}, which is similar to a Gaussian function:
\begin{equation}
\frac{dN}{dz}(z)=\frac{1}{\sqrt{2\pi}\,\sigma}\exp{\left[-\frac{(z-z_0)^2}{2\,\sigma^2}\right]}~,
\end{equation}
where the mean redshift $z_0=0.55$ and the deviation $\sigma=0.055$.

\subsection{The Power Spectrum Measurement}

\subsubsection{The pseudo-$C_\ell$ estimator}

A two dimensional density field $\sigma({\bf n})$ defined over the
full sky can be decomposed in a series of spherical harmonics
$Y_{\ell m}$ and their corresponding coefficients ${a}_{\ell m}$:
\begin{equation}
\sigma({\bf n})=\sum^\infty_{\ell=0}\sum^{\ell}_{m=-\ell}{a}_{\ell
m}Y_{\ell m}({\bf n})~,
\end{equation}
with
\begin{equation}
{a}_{\ell m}=\int d\Omega_{\bf n}\,\sigma({\bf n}) Y^\ast_{\ell
m}({\bf n})~.
\end{equation}
On the full sky spherical harmonics are a complete and orthogonal
basis set. However, in practice,  a masked  (incomplete) sky is
observed. Thus the observed $\tilde{a}_{\ell m}$ coefficients on the
partially cut sky become \cite{Hivon}:
\begin{equation}
\tilde{a}_{\ell m}=\int d\Omega_{\bf n}\,\sigma({\bf n})W({\bf n})
Y^\ast_{\ell,m}({\bf n})~,
\end{equation}
where $W({\bf n})$ is the position dependent,  weight function
imposed by the mask and, optionally, the adopted weighting scheme.
From these coefficients we can obtain the observed pseudo power
spectrum $\tilde{C}_\ell$, the pseudo-$C_\ell$ estimator
\cite{Hivon}:
\begin{equation}
\tilde{C}_\ell=\frac{1}{2\ell+1}\sum^\ell_{m=\ell}\left|\tilde{a}_{\ell
m}\right|^2~.
\end{equation}
The pseudo power spectrum $\tilde{C}_\ell$, given by the weighted
spherical harmonic transform of a  masked map, is clearly different
from the full sky true power spectrum ${C}_\ell$. The expectation
value of $\tilde{C}_\ell$ is related to the true ${C}_\ell$ by a
convolution:
\begin{equation}
\langle\tilde{C}_\ell\rangle=\sum_{\ell'}{C}_{\ell'}G_{\ell\ell'}~,
\end{equation}
where the coupling matrix $G_{\ell\ell'}$, describing the mode
coupling resulting from the weight function $W({\bf n})$
\cite{peeblesGll}, can be expressed in terms of $3j$ symbols as:
\begin{eqnarray}
G_{\ell_1\ell_2}=(2\ell_2+1)\sum_{\ell_3=|\ell_1-\ell_2|}^{\ell_1+\ell_2}
\frac{2\ell_3+1}{4\pi}\tilde{\mathcal{W}}_{\ell_3}\left(
\begin{array}{ccc}
\ell_1&\ell_2&\ell_3\\
0&0&0
\end{array}\right)^2~.\label{eq:gll}
\end{eqnarray}
$\tilde{\mathcal{W}}_{\ell}$ is the angular power spectrum of the
weight function:
\begin{equation}
\tilde{\mathcal{W}}_\ell=\frac{1}{2\ell+1}\sum^\ell_{m=\ell}\left|\tilde{w}_{\ell
m}\right|^2~,
\end{equation}
where the weight coefficients $\tilde{w}_{\ell m}$ are:
\begin{equation}
\tilde{w}_{\ell m}=\int d\Omega_{\bf n}W({\bf n})
Y^\ast_{\ell,m}({\bf n})~.
\end{equation}

For the LSS catalogues presented above, in practice, we construct
the pixelized maps using the HEALPix software package \cite{healpix}
with high resolution $N_{\rm side} = 512$, yielding pixel areas of
$6.87'\times 6.87'$ and estimate their pseudo power spectrum
$\hat{C}_\ell$ using the HEALPix function ``{\tt anafast}''.  We
also subtract the shot noise contribution $\Delta\Omega/N$ from the
observed pseudo power spectrum, where $\Delta\Omega$ is the surveyed
area and $N$ is the observed number of sources. The pseudo power
spectrum is measured up to $\ell_{\rm max}=400$, at these multipoles
the magnitude of the estimated error bars is dominated by
shot-noise. In order to avoid the effects of gauge corrections on
the power spectrum on very large scales \cite{Yoo1,Yoo2}, we set the
minimal multipole $\ell_{\rm min}=10$. The effects of
non-Gaussianity are large on very large scales. By ignoring
$\ell<10$ we are neglecting the scales where the signal is maximal.
However the theoretical predictions for the measured correlations is
uncertain on very large scales as discussed in ref. \cite{Yoo2}. In
ref. \cite{xiaCCF} we estimated that these uncertainties could
introduce a systematic error on $f_{\rm NL}$ of $5$. Here we take a
more conservative approach and simply ignore the largest scales,
even if this implies a reduction of the signal-to-noise.

\subsubsection{Covariance matrix of the pseudo-$C_\ell$ estimator}

In the signal dominated limit, the inverse of the power spectrum
covariance matrix $F_{\ell\ell'}$ is \cite{Hinshaw}:
\begin{equation}
F^{\rm
mask}_{\ell\ell'}=\frac{(2\ell+1)G_{\ell\ell'}}{2(C_\ell+N_\ell)(C_{\ell'}+N_{\ell'})}~,
\end{equation}
where $C_\ell$ is the theoretical angular power spectrum,
$N_\ell=\Delta\Omega/N$ denotes the shot noise term. $C_\ell$ and
$N_\ell$ are the necessary error contributions from both cosmic
variance and shot noise, respectively. For the full sky survey,
$G_{\ell_1\ell_2}=\delta_{\ell_1\ell_2}f_{\rm sky}$ and the
covariance matrix becomes diagonal:
\begin{equation}
F^{\rm full}_{\ell\ell'}=\frac{(2\ell+1)f_{\rm
sky}}{2(C_\ell+N_\ell)(C_{\ell'}+N_{\ell'})} \delta_{\ell \ell'}~.
\end{equation}

\subsubsection{Cross-correlation power spectrum}

Besides the angular power spectrum of each LSS tracer, we also
consider the cross-correlation power spectrum among them. In figure
\ref{fig:redshift}, we show that the redshift distributions of these
three tracers overlap. We can thus expect some cross-correlation
signal among them. We also use the HEALPix software to estimate the
cross-correlation power spectrum $\tilde{C}_\ell^{\rm XY}$ between
tracer $X$ and $Y$:
\begin{equation}
\tilde{C}_\ell^{\rm
XY}=\frac{1}{2\ell+1}\sum^\ell_{m=\ell}\tilde{a}^{\rm X}_{\ell
m}\tilde{a}^{\rm \ast Y}_{\ell m}~,
\end{equation}
where $\tilde{a}^{\rm i}_{\ell m}$ are the observed coefficients of
tracer ${\rm i}$ and they are computed on the common sky area only.
Consequently, the inverse of the covariance matrix becomes:
\begin{equation}
F^{\rm
mask}_{\ell\ell'}=\frac{(2\ell+1)G_{\ell\ell'}}{\sqrt{(C_\ell^{\rm
XY})^2+ (C_\ell^{\rm XX}+N_\ell^{\rm XX})(C_\ell^{\rm
YY}+N_\ell^{\rm YY})}\sqrt{(C_{\ell'}^{\rm XY})^2+(C_{\ell'}^{\rm
XX}+N_{\ell'}^{\rm XX})(C_{\ell'}^{\rm YY}+N_{\ell'}^{\rm YY})}}~,
\end{equation}
where the shot noise term $N_\ell^{\rm i}=\Delta\Omega/N_{\rm i}$ is
scale independent, $\Delta\Omega$ is the common area of two surveys,
$N_{\rm i}$ is the observed number of sources in this common area.

Similarly, we also estimate the cross-correlation power spectrum
$\tilde{C}_{\ell}^{\rm XT}$ between the LSS number density map and
the CMB map. While the Planck satellite is operating \cite{planck},
here, we use the the 7-years Internal Linear Combination (ILC) map
of the CMB provided by the Wilkinson Microwave Anisotropy Probe
(WMAP) team \cite{ilc7}. Since an estimate of the uncertainty
associated to this map is not provided, we have checked that using
other CMB maps (e.g. the 5 years ILC map by the WMAP team) does not
change the results in any significant way. For the cross-check, we
also use the V and W band all-sky 7-years CMB maps provided by WMAP
\cite{Jarosik11} and smooth these maps to the resolution of ILC map
(60 arcmins). Using these maps, we perform all the calculations
again and find that our results are undistinguishable. We adopt the
WMAP KQ75 mask, excluding about 30\% of the sky at low Galactic
latitude, to avoid most of the residual Galactic contamination.

We still use the estimator of cross-correlation power spectrum
$\tilde{C}_\ell^{\rm XT}$:
\begin{equation}
\tilde{C}_\ell^{\rm
XT}=\frac{1}{2\ell+1}\sum^\ell_{m=\ell}\tilde{a}^{\rm X}_{\ell
m}\tilde{a}^{\rm \ast T}_{\ell m}~,
\end{equation}
where $\tilde{a}^{\rm T}_{\ell m}$ are the observed coefficients of
the CMB map $\Delta T({\bf n})$, also computed on the common sky
area only. The inverse of the covariance matrix is:
\begin{equation}
F^{\rm
mask}_{\ell\ell'}=\frac{(2\ell+1)G_{\ell\ell'}}{\sqrt{(C_\ell^{\rm
XT})^2+ (C_\ell^{\rm XX}+N_\ell^{\rm XX})(C_\ell^{\rm
TT}+N_\ell^{\rm TT})}\sqrt{(C_{\ell'}^{\rm XT})^2+(C_{\ell'}^{\rm
XX}+N_{\ell'}^{\rm XX})(C_{\ell'}^{\rm TT}+N_{\ell'}^{\rm TT})}}~,
\end{equation}
where $C_\ell^{\rm TT}$ is the theoretical angular power spectrum of
CMB map, $N_\ell^{\rm TT}$ is measurement error of the CMB
temperature power spectrum, in which the main source is the pixel
noise which accounts for the imperfections in measured temperature
on the sky. Pixel noise for WMAP is reported as $\sigma_{\rm
pix}=\sigma_0 / \sqrt{N_{\rm obs}}$, where $\sigma_0$ is the noise
per observation and $N_{\rm obs}$ is the number of observations per
given pixel \cite{nobs}. After getting the theoretical $C_\ell^{\rm
T}$ and $C_\ell^{\rm XT}$, we include the correction of Gaussian
beam $B(\ell)$ (using the HEALpix {\tt gaussbeam} routine):
$C_\ell^{\rm TT}=C_\ell^{\rm TT}B(\ell)^2$, $C_\ell^{\rm
XT}=C_\ell^{\rm XT}B(\ell)$. We neglect the pixel window function
since its contribution at $N_{\rm side}=512$ is negligible at the
resolution of the WMAP ILC map which is $60$ arcmins FWHM.

\section{Method \&  Data Analysis}\label{method}

We use the publicly available package {\tt CAMB$_{-}$sources}
\footnote{Available at http://camb.info/sources/.} \cite{camb} to
calculate the theoretical power spectrum. In our analysis, we use
the Halofit \cite{Halofit} built-in routine for non-linear
correction to obtain the fully-evolved, nonlinear matter power
spectrum $P(k)$ at any epoch. We perform a global fitting of
cosmological parameters, including $f_{\rm NL}$, for the data of
section \ref{data} including also the datasets described below,
using the {\tt CosmoMC} package \cite{cosmomc}, a Markov Chain Monte
Carlo (MCMC) code, modified to calculate the theoretical ACF and
CCF. We assume purely adiabatic initial conditions and a flat
Universe, with no tensor contribution to primordial fluctuations.
The following six cosmological parameters are allowed to vary with
top-hat priors: the dark matter energy density parameter
$\Omega_{\rm c} h^2 \in [0.01,0.99]$, the baryon energy density
parameter $\Omega_{\rm b} h^2 \in [0.005,0.1]$, the primordial
spectral index $n_{\rm s} \in [0.5,1.5]$, the primordial amplitude
$\log[10^{10} A_{\rm s}] \in [2.7,4.0]$, the ratio (multiplied by
100) of the sound horizon at decoupling to the angular diameter
distance to the last scattering surface $\Theta_{\rm s} \in
[0.5,10]$, and the optical depth to reionization $\tau \in
[0.01,0.8]$. The pivot scale is set at $k_{\rm s0}=0.05\,$Mpc$^{-1}$
and do not consider massive neutrinos and dynamical dark energy.
Besides these six basic cosmological parameters, we have two more
parameters related to the power spectrum $C_\ell$ data: the
non-Gaussianity parameter, the minimal halo mass $M_{\rm min}$. When
we combine all the different probes together we have four
parameters: three minimal halo masses for three LSS surveys and the
non-Gaussianity parameter.

The model power spectrum $C_\ell^{\rm th}$ is compared with the
observed values $\tilde{C}_\ell$, respectively, through the Gaussian
likelihood function \cite{liciachi2}:
\begin{eqnarray}
\ln\mathcal{L}\propto-\frac{1}{2}\sum_{\ell\ell'}(C_\ell^{\rm
th}-\tilde{C}_\ell){F_{\ell\ell'}(C_{\ell'}^{\rm
th}-\tilde{C}_{\ell'})}~,
\end{eqnarray}
where the Fisher matrix $F_{\ell\ell'}$ is the inverse of the power
spectrum covariance matrix. While it is true that at very low $\ell$ a
Gaussian likelihood is not a good approximation for CMB data, at high
$\ell$ the Gaussian approximation works well even for CMB, for LSS it
is customary to adopt a Gaussian likelihood.

The following cosmological data are also included in the fit: ${\rm
i})$ power spectra of CMB temperature and polarization anisotropies;
${\rm ii})$ baryonic acoustic oscillations (BAOs) in the galaxy
power spectra; ${\rm iii})$ SNIa distance moduli.

To deal with the 7-years WMAP (WMAP7) CMB temperature and polarization
power spectra we use the routines for computing the likelihood
supplied by the WMAP team \cite{WMAP7}. The WMAP7 data are used only
to improve the constraints on the six basic cosmological parameters,
not to constrain $f_{\rm NL}$.

The BAOs \cite{BAO} can, in principle, measure not only the angular
diameter distance, $D_A(z)$, but also the expansion rate of the
Universe, $H(z)$. However, the limited accuracy of current data only
allows us to determine the ratio between the distance scale defined
by ref. \cite{Eisenstein:2005su}:
\begin{equation}
D_v(z)\equiv\left[(1+z)^2D_A^2(z)\frac{cz}{H(z)}\right]^{1/3},
\end{equation}
and the comoving sound horizon at the baryon-drag epoch, $r_s(z_d)$
(see ref. \cite{Eisenstein:1997ik}). Accurate determinations of the
distance ratio $r_s(z_d)/D_v(z)$, $r_s(z_d)$ have been obtained by
Ref. \cite{BAO}:
\begin{eqnarray}
r_s(z_d)/D_v(z=0.20)&=&0.1905\pm0.0061,\nonumber\\
r_s(z_d)/D_v(z=0.35)&=&0.1097\pm0.0036.
\end{eqnarray}
We adopt these values as Gaussian priors.

The SNIa distance moduli provide the luminosity distance as a
function of redshift $D_{\rm L}(z)$ which provides strong
constraints on the dark energy evolution. In this paper we will use
the latest SNIa data sets from the Supernova Cosmology Project,
``Union2 Compilation'' which consists of 557 samples and spans the
redshift range $0\lesssim z \lesssim 1.55$ \cite{Amanullah:2010vv}.
In this data set, they improved the data analysis method by using
and refining the approach of their previous work
\cite{Kowalski:2008ez}. When comparing with the previous ``Union
Compilation'', they extended the sample with the supernovae from
refs. \cite{Amanullah:2010vv,Amanullah:2007yv,Holtzman08,Hicken09}.
The authors also provide the covariance matrix of data with and
without systematic errors and, in order to be conservative, we
include systematic errors in our calculations. In the calculation of
the likelihood from SNIa we marginalize over the nuisance parameter
as done in refs. \cite{SNMethod1,SNMethod2}.

Furthermore, we add a prior on the Hubble constant, $H_0=74.2 \pm
3.6$ km/s/Mpc given by ref. \cite{HST}. Finally, in the analyses we
set the minimal halo mass at $M_{\rm min} > 10^{12} h^{-1}M_{\odot}$
consistent with the results of ref. \cite{Croom}.

%%%%%%%%%%%%%%%%%%%%%%%%%%%%%%%%%%%%

\begin{figure*}[htpb]
\begin{center}
\includegraphics[scale=0.3]{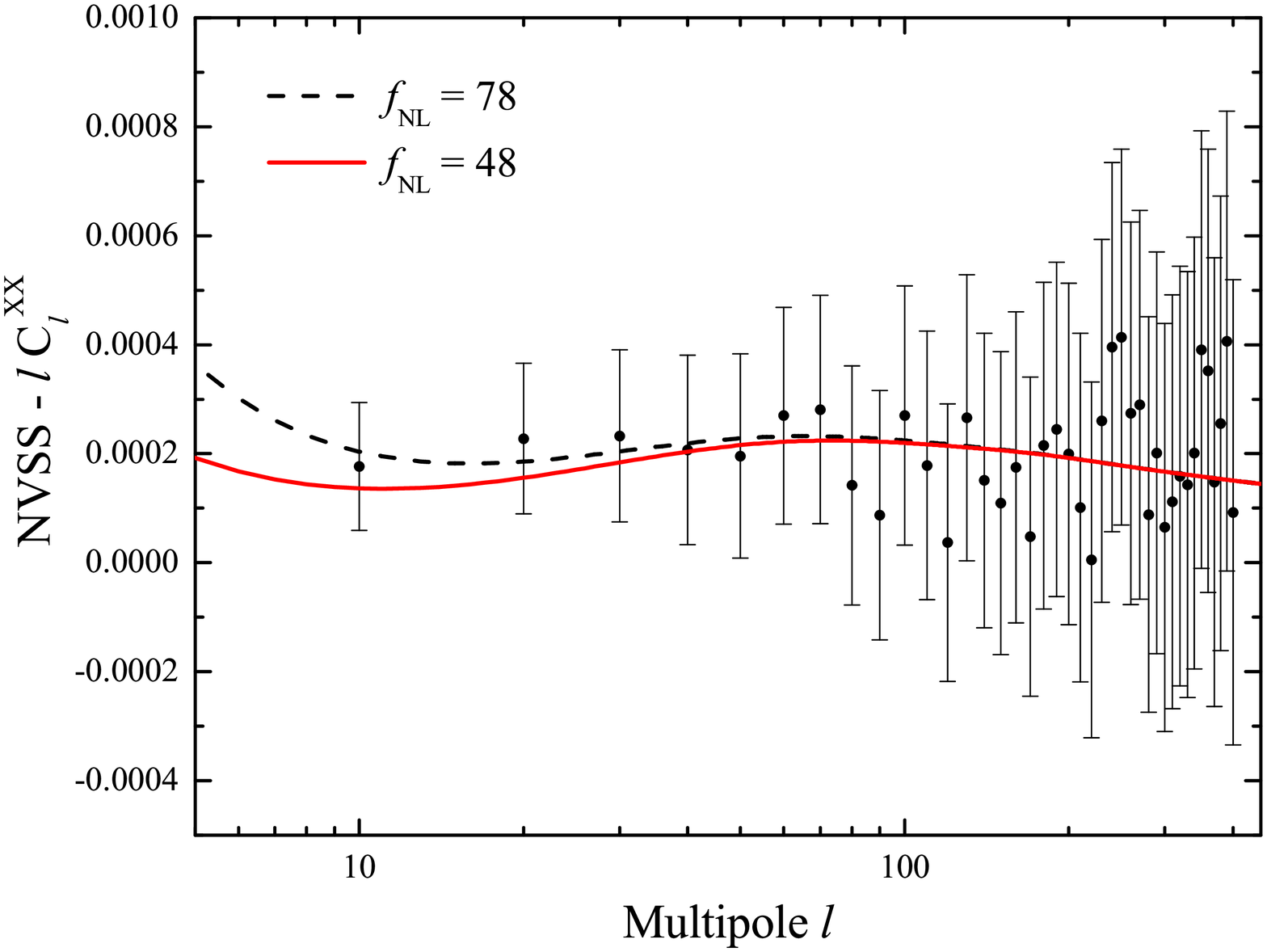}
\includegraphics[scale=0.3]{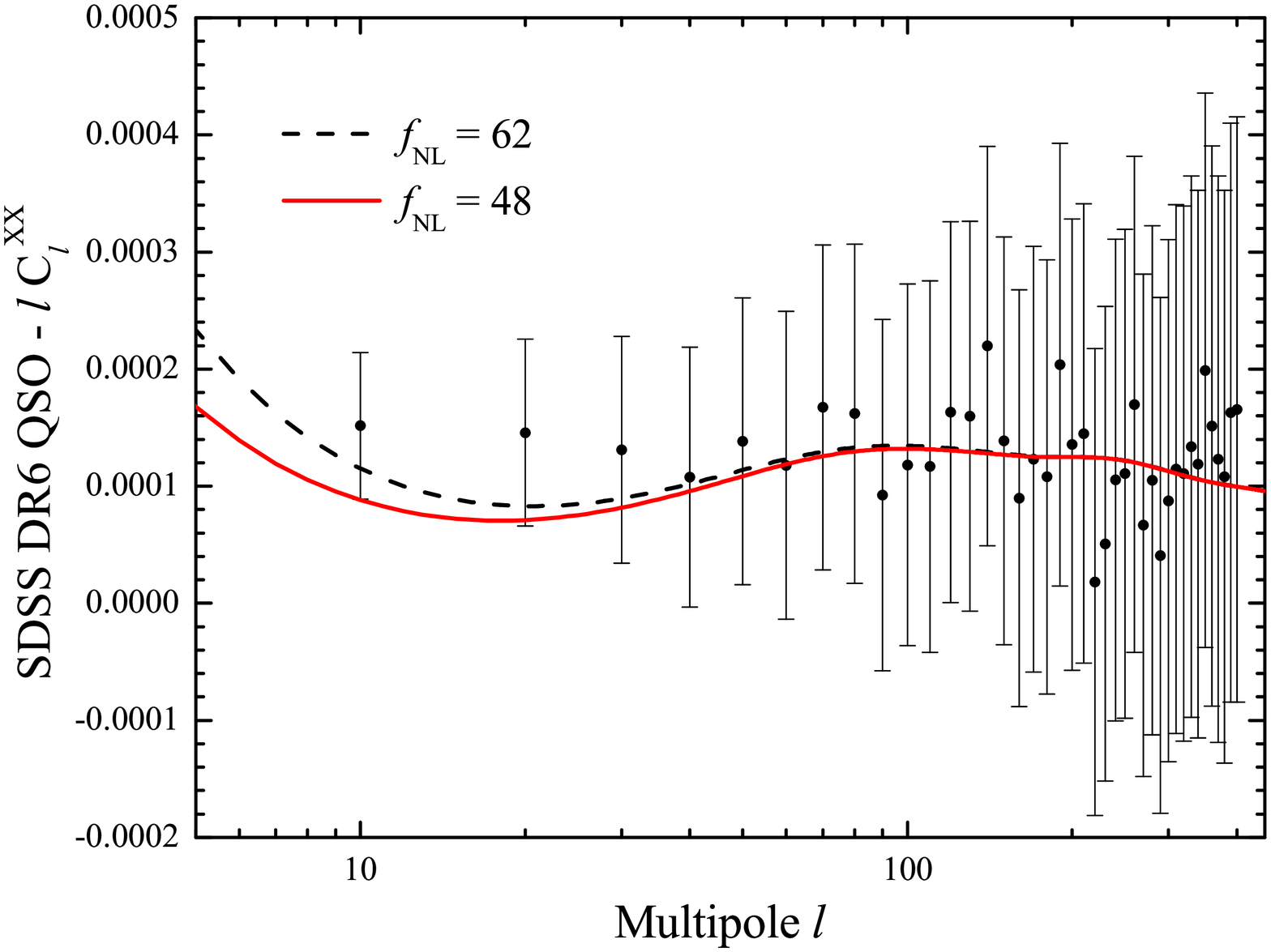}
\includegraphics[scale=0.3]{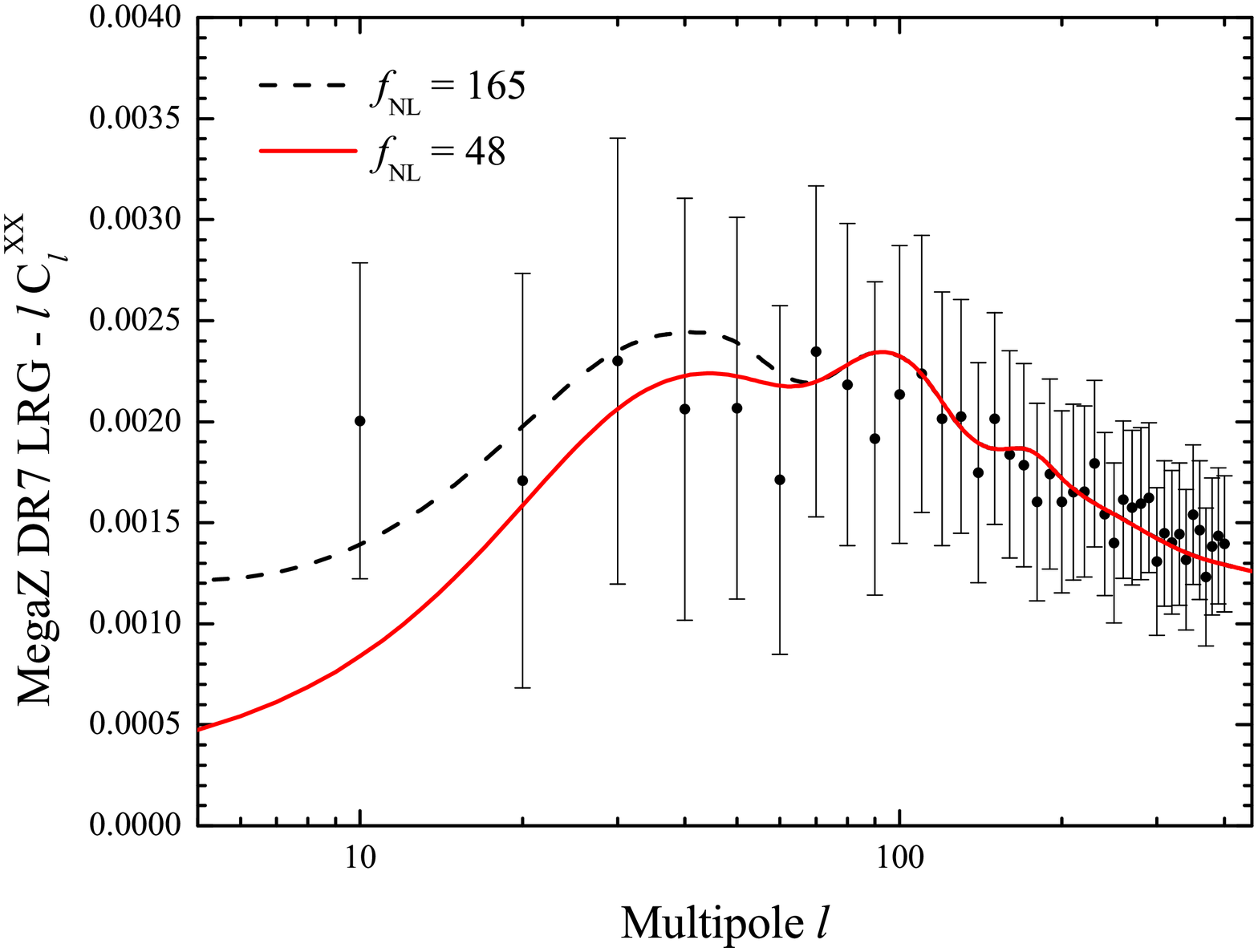}
\caption{Observed angular power spectra of three LSS tracers. The
black dashed lines are the best fit models using angular power
spectra only. The red solid lines are the best fit models $f_{\rm
NL}=48$ when using all data together. For illustration purposes, we
show the binned power spectra with the bin size $\Delta\ell=10$.
\label{fig:data1}}
\end{center}
\end{figure*}

\begin{figure*}[htpb]
\begin{center}
\includegraphics[scale=0.28]{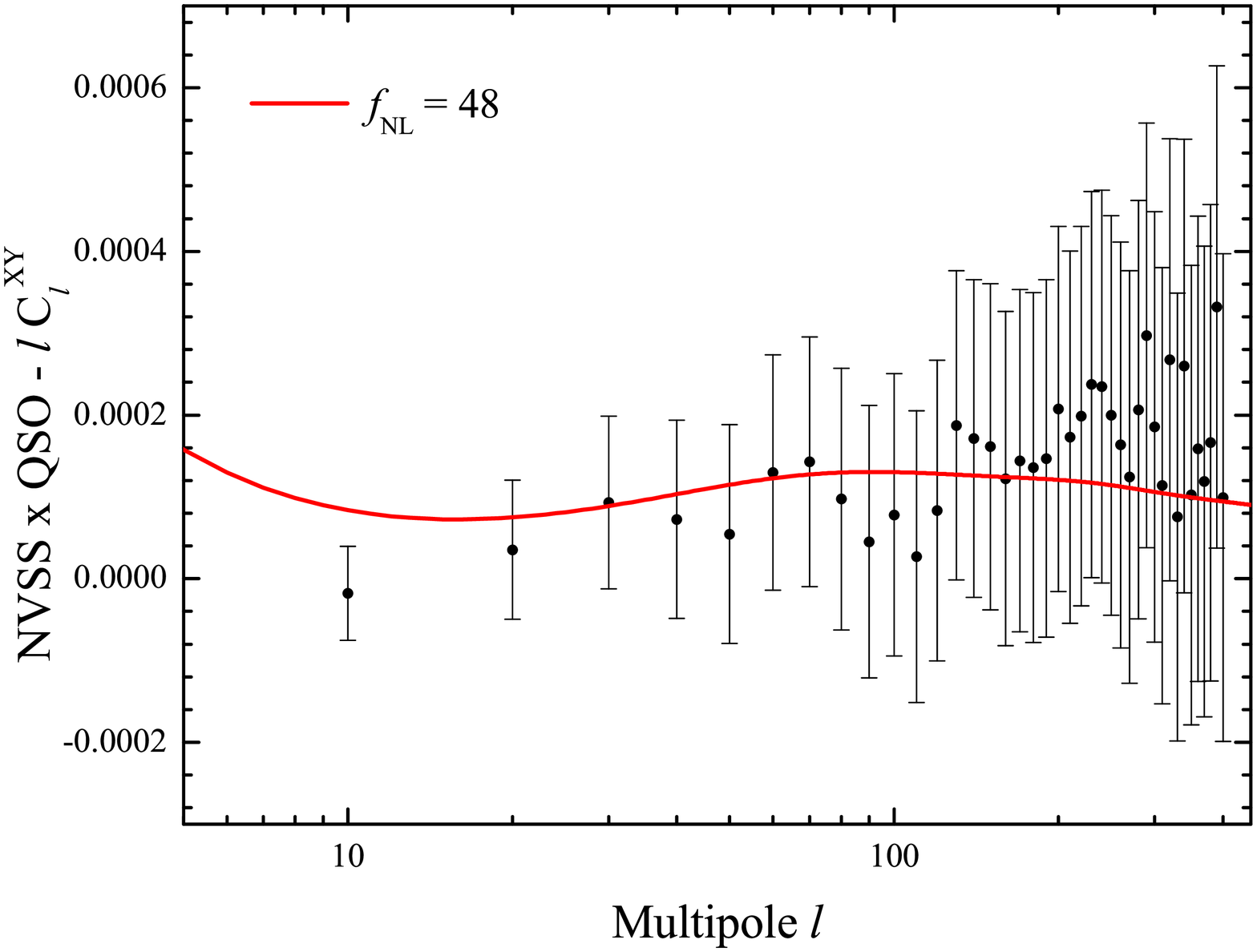}
\includegraphics[scale=0.28]{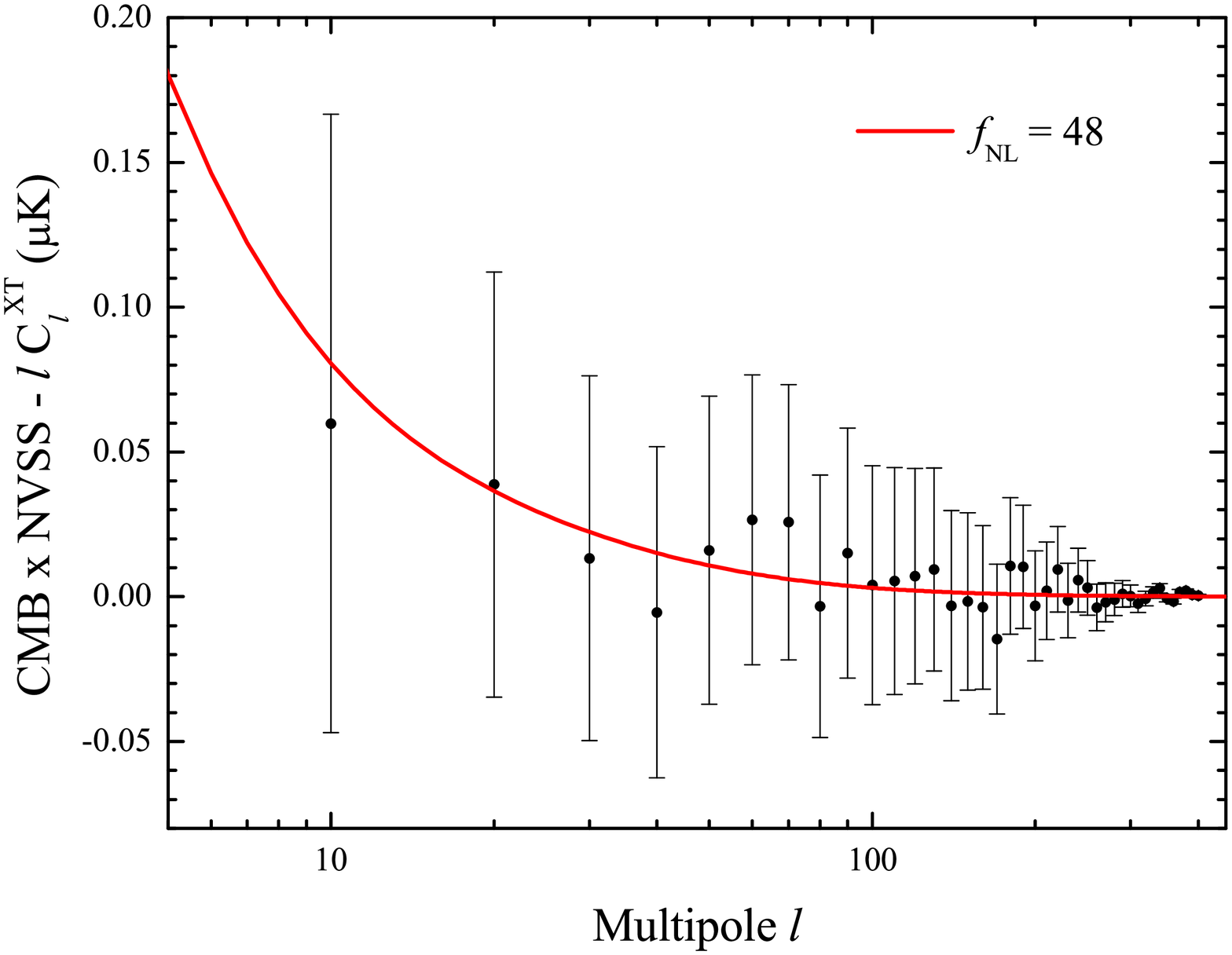}
\includegraphics[scale=0.28]{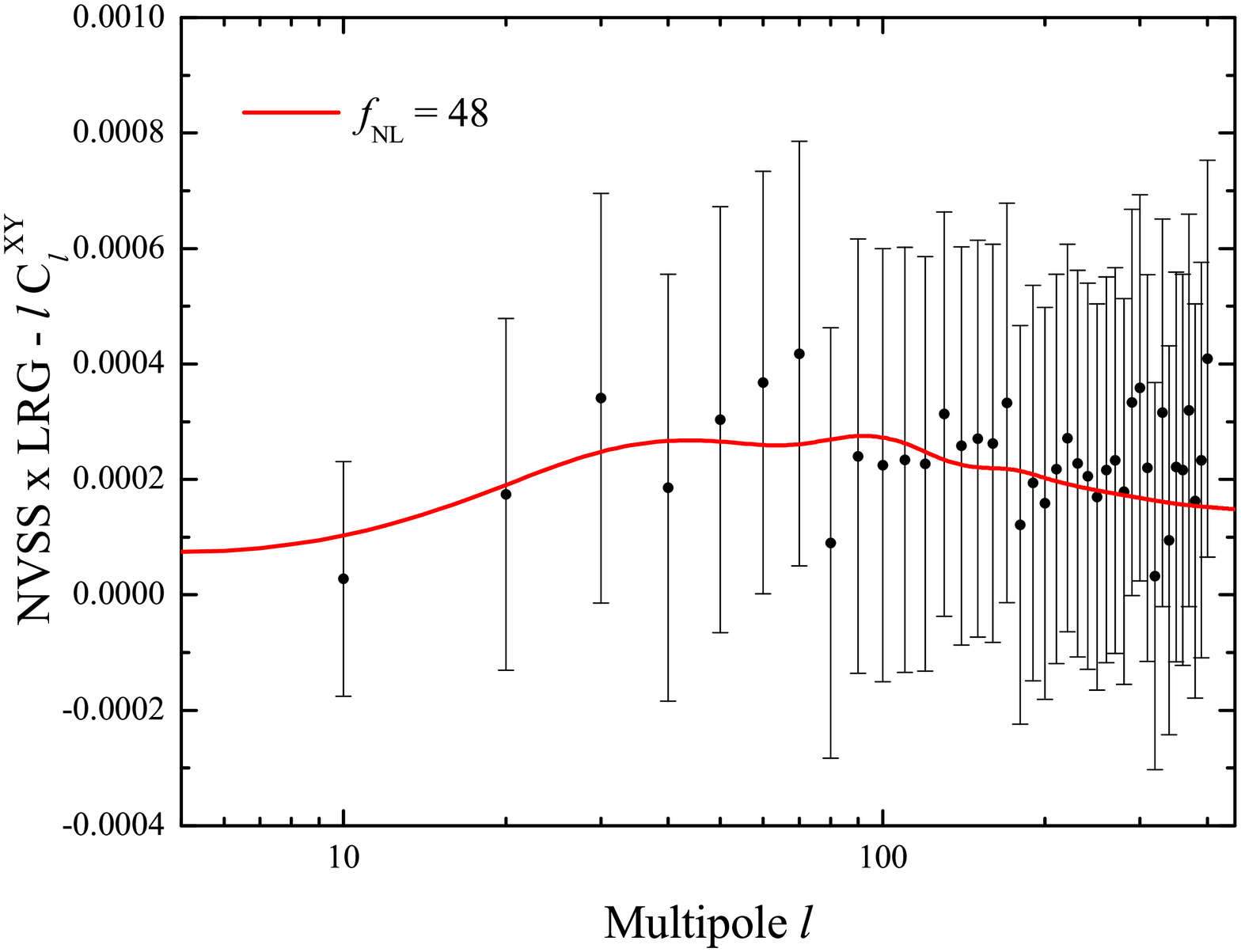}
\includegraphics[scale=0.28]{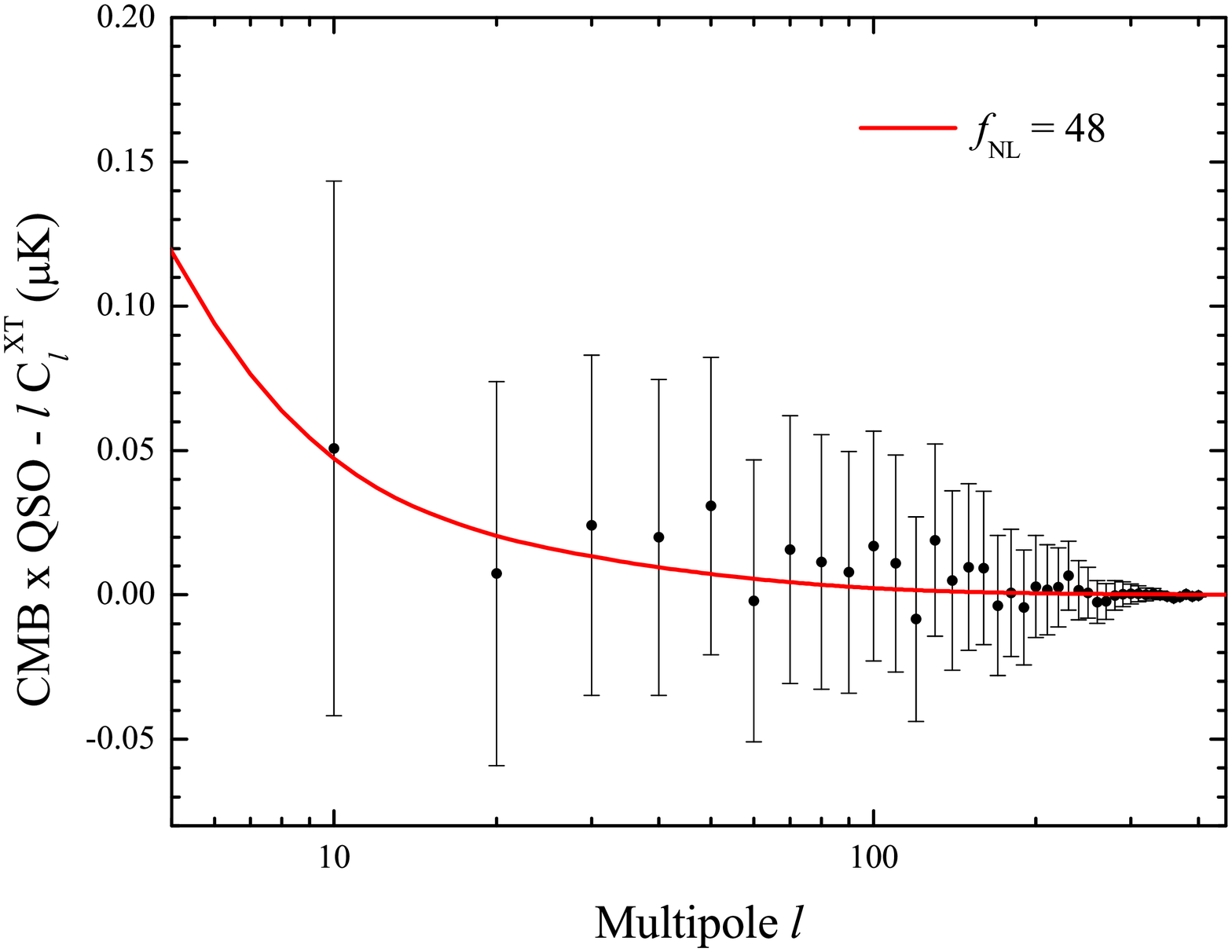}
\includegraphics[scale=0.28]{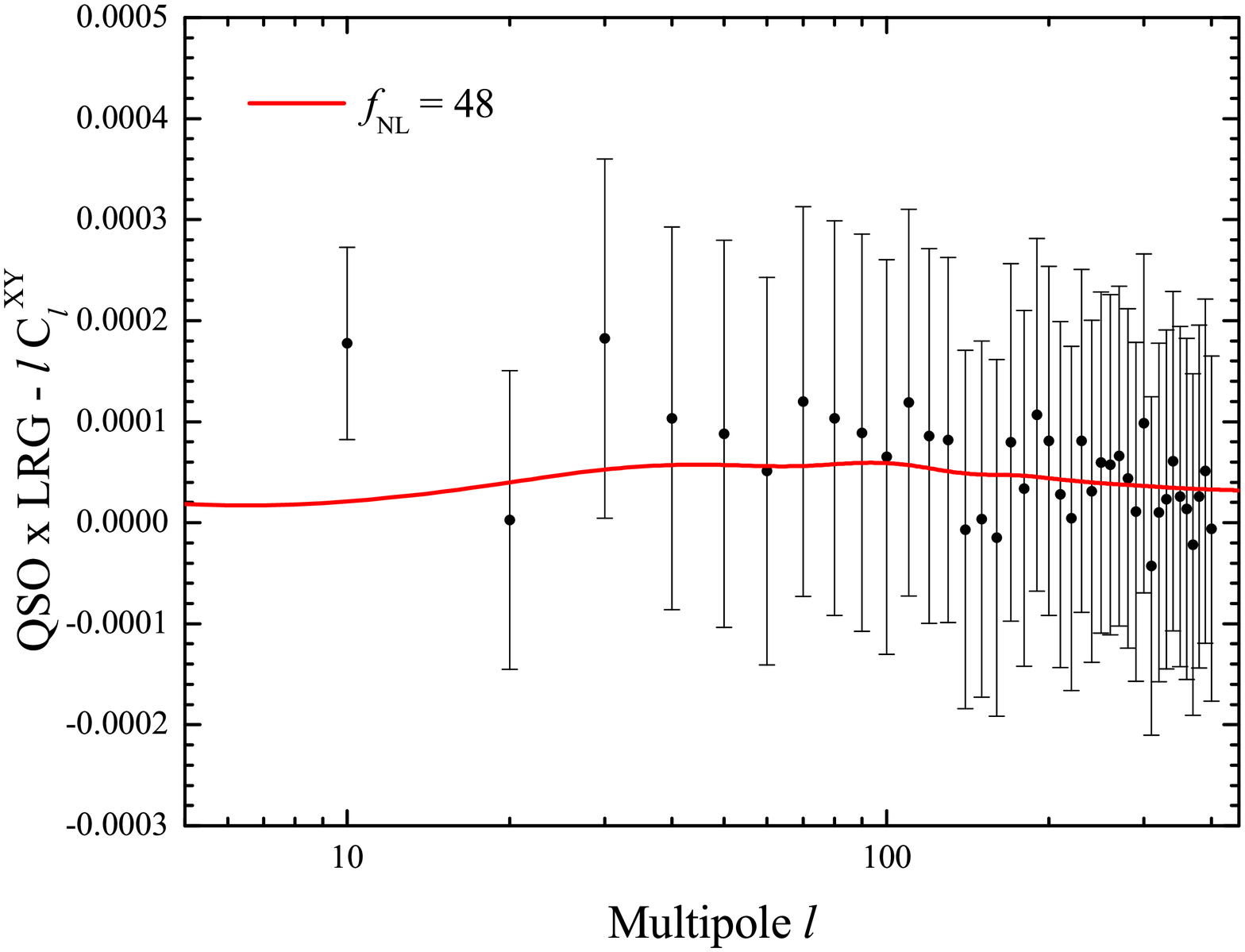}
\includegraphics[scale=0.28]{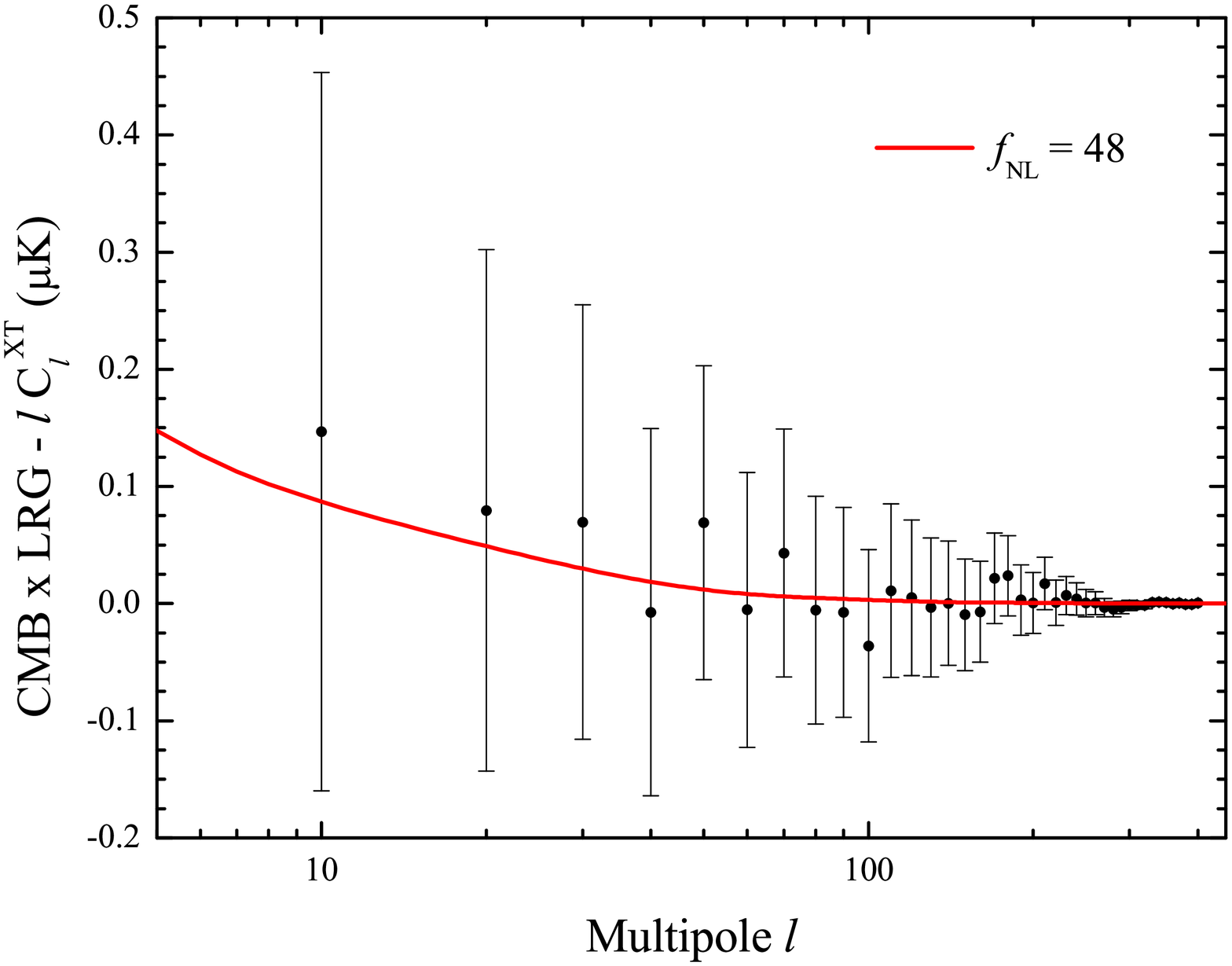}
\caption{Left Panels: Observed cross-correlation power spectra among
three LSS tracers. Right Panels: Observed cross-correlation power
spectra between CMB ILC map and three LSS tracers. The red solid
lines are the best fit models $f_{\rm NL}=48$ when using all data
together. For illustration purposes, we show the binned power
spectra with the bin size $\Delta\ell=10$. \label{fig:data2}}
\end{center}
\end{figure*}

\section{Numerical Results}\label{result}

In this section, we present constraints on the primordial
non-Gaussianity  parameter $f_{\rm NL}$ from the observed  LSS power
spectra, in combination with the external data sets introduced
above: the WMAP7, BAO and SNIa. These extra datasets are only used
to constrain the underlying cosmology. In table \ref{tab:I} and
figure \ref{fig:fnl} we show the constraints on the primordial
non-Gaussianity parameter from different data combinations, after
marginalizing over all the other parameters. We also plot the
observed power spectra data $\ell \tilde{C}_\ell$ in figure
\ref{fig:data1} and figure \ref{fig:data2}, together with the
corresponding theoretical best fit models for comparison. For
illustration purposes, here we bin the observed power spectra with
the bin size $\Delta\ell=10$.

\subsection{Local Type}

\subsubsection{Angular Power Spectrum}\label{resNVSS}

We begin by  considering the constraint on the local type of
non-Gaussianity $f_{\rm NL}$ from the angular power spectrum only.
In the first panel of figure \ref{fig:data1}, we plot the observed
NVSS angular power spectrum data, which is consistent with the
previous work \cite{blakeNVSS}. When using the NVSS $\tilde{C}_\ell$
data only, we obtain the marginalized constraint on the
non-Gaussianity parameter: $f_{\rm NL}=78\pm52$ ($1\,\sigma$) and
$-34 < f_{\rm NL} < 187$ ($2\,\sigma$), which is consistent with
zero.  The minimal halo mass, $M_{\rm
min}=10^{12.44\pm0.26}h^{-1}M_{\odot}$ ($1\,\sigma$), turns out to
be remarkably close to our previous work \cite{xiaACF}. We show the
best fit model in this panel (black dashed line). The amplitude of
the power spectrum on large scales (small $\ell$) is enhanced to fit
the data points, by the positive value of $f_{\rm NL}$. In our
previous works \cite{xiaACF,xiaCCF}, we used the auto-correlation
function (ACF) of NVSS radio sources to constrain $f_{\rm NL}$ and
found that the NVSS ACF yields a positive $f_{\rm NL}$ at more than
$2\,\sigma$. The reason for this difference is that we only use the
power spectrum data at $\ell \geq 10$, and the power spectrum
approach does not include the information on the total number of
NVSS radio sources $N_{\rm tot}$ (which was folded in the ACF
analysis). If we add a prior on the total number of sources $N_{\rm
tot}$: $0.5 < {N_{\rm tot}}/{N_{\rm tot}^{\rm NVSS}} < 1.5$, we
obtain a limit of $f_{\rm NL}=62\pm30$ at $1\,\sigma$. This result
is perfectly consistent with previous works \cite{xiaACF,xiaCCF}.

Next, we use the estimated angular power spectrum of the SDSS DR6
QSO sample, which is shown in the second panel of figure
\ref{fig:data1}. There is an obvious excess power at large scales,
which favors a positive $f_{\rm NL}$: $f_{\rm NL}=62\pm26$
($1\,\sigma$) and $5 < f_{\rm NL} < 115$ ($2\,\sigma$). Similarly to
our previous work \cite{xiaCCF}, we also find evidence for positive
$f_{\rm NL}$ at more than $2\,\sigma$. However, ref.~\cite{Slosar},
using SDSS QSO data, obtained non-Gaussianity constraints consistent
with zero at 95\% confidence level. As we discussed before,
ref.~\cite{Slosar} used an extension of the DR3 QSO sample that
include sources that subsequently were released with the DR6 sample:
we use the more complete and better calibrated final official SDSS
DR6 QSO catalog release \cite{richards09}. Ref. \cite{Slosar} found
that their QSO sample at $z < 1.45$ (QSO0) seem to suffer from
contamination which they attribute to systematic calibration errors,
discarded this sample from their analysis and only used the high
redshift sample (QSO1). We have checked for this effect and find no
evidence in the sample we use that the $z < 1.45$ QSOs have
different contamination than the $z > 1.45$ \cite{xiaCCF}. In
addition, we remove three narrow stripes in the southern hemisphere
of the galaxy and set the minimal multipole $\ell_{\rm min}=10$,
which could reduce the effect of star contamination further.

Finally, we consider the MegaZ DR7 LRG sample. Ref. \cite{LRG7}
split this sample in four redshift bins according to each source
photometric redshift estimate and found a large excess of power over
the lowest multipoles ($\ell < 20$) in the angular power spectrum,
which is also found in their previous analysis using MegaZ DR4 LRG
sample \cite{blakeLRG}. In our analysis, although we do not split
the LRG sample in redshift bins, we still find this excess power on
large scales, which is shown in the third panel of figure
\ref{fig:data1}. There are several explanations for this excess
power that range from systematic errors as due to contaminants to
new physics \cite{LRGPRL}. Ref. \cite{LRGPRL} checked some possible
systematic errors of this MegaZ LRG sample in details and found that
systematics errors do not seem to be responsible for the excess
power at large scales. We also perform a similar check and find no
significant changes in the power spectrum at large scales. One of
the possibilities is that this excess power could be induced by the
primordial non-Gaussianity. Here, we use the observed
$\tilde{C}_\ell$ data of MegaZ DR7 LRG sample to constrain the
primordial non-Gaussianity for the first time. We calculate the
angular power spectrum on largest scales and account for the
redshift distortion power as described in Ref.
\cite{PadmanabhanLRG}, although we set the minimal multipole
$\ell_{\rm min}=10$. After marginalizing over other free parameters,
we obtain the constraint:
\begin{equation}
f_{\rm NL}=165\pm105~~(1\,\sigma)~,~~~~-81 < f_{\rm NL} <
351~~(2\,\sigma)~,
\end{equation}
which is consistent with zero at $2\,\sigma$. We also obtain a limit
of $M_{\rm min}=10^{12.45\pm0.07}h^{-1}M_{\odot}$ at 68\% confidence
level. Using eq. (\ref{eq:nghalobias}), we could obtain the limit of
effective halo bias at the mean redshift of the survey:
$b(z=0.55)=1.93\pm0.06$ ($1\,\sigma$), which is consistent with
other works \cite{Fantaye}, although in our analysis we directly use
the full sample.

When we combine these three observed angular power spectra data
together to constrain non-Gaussianity, we obtain a limit of $f_{\rm
NL}=68\pm22$ at 68\% confidence level and $22<f_{\rm NL}<108$ at
95\% confidence level, which implies the current angular power
spectra of LSS tracers favor $f_{\rm NL}>0$ at the $\sim 3\,\sigma$
level, since non-Gaussianity adds power on large angular scales
yielding a good fit to the observed data points.

As can be seen from figure \ref{fig:data1} and table \ref{tab:I},
there is not a specific measurement that is driving the signal: all
the different probes yield  very consistent results and  when
combined the error bar on $f_{\rm NL}$ is reduced to yield  a $>
2\sigma$ result.

\subsubsection{Adding Cross-Correlation Power Spectrum}

Comparing with the angular power spectrum, the constraining power of
cross-correlation power spectrum on $f_{\rm NL}$ is much  weaker
\cite{melita}. In the panels of figure \ref{fig:data2}, we plot the
observed cross-correlation power spectra $\tilde{C}_\ell^{\rm XY}$
among three LSS tracers and $\tilde{C}_\ell^{\rm XT}$ between the
CMB map and three LSS tracers. These cross-correlation data points
are consistent with zero within $1\,\sigma$ error bar but the best
fit value $f_{\rm NL}=48$ is an excellent fit to the data points, as
shown in figure \ref{fig:data2}. However, cross-correlation data
give a larger error bar for $f_{\rm NL}$ than that from the angular
power spectra data and are thus much less constraining.

First, we consider the single tracer case in combination with its
cross-correlation with the CMB map. When we combine the angular
power spectrum and its cross-correlation power spectrum
$\tilde{C}_\ell^{\rm XT}$ with CMB map to constrain the primordial
non-Gaussianity, we obtain $f_{\rm NL}=74\pm40$, $f_{\rm
NL}=59\pm21$ and $f_{\rm NL}=153\pm95$ at 68\% confidence level for
NVSS radio sources, SDSS DR6 QSOs and MegaZ DR7 LRGs, respectively.
The results are consistent with those obtained from the angular
power spectrum only, but those cross correlations does not seem to
add much information.

If we consider all the angular and cross-correlation power spectra
data sets of three LSS tracers together \footnote{In our analysis,
we do not take into account the possible correlation between the
auto-correlation and cross-correlation power spectra.  Neglecting
this correlation should have a  negligible effect. In fact  when the
cross-correlation signal is much weaker than the auto-correlation
signal, as it happens here, their covariance is diagonal-dominated
with negligible off-diagonal terms.  This is well known to happen
for example when considering the ISW effect  from the galaxy-CMB
cross correlation  and the CMB temperature $C_{\ell}$.}, we obtain
the constraint on the local type primordial non-Gaussianity:
\begin{equation}
f_{\rm NL}=48\pm20~~(1\,\sigma)~,~~~~5 < f_{\rm NL} <
84~~(2\,\sigma)~.
\end{equation}
This result is compatible with previous estimates
\cite{Slosar,Yadav:2007yy,Pietrobon:2008ve,Curto:2009pv,Smidt:2009ir,JV,Smith:2009jr,Rud},
and with the WMAP7 limits \cite{WMAP7}. The LSS data still implies a
positive $f_{\rm NL}$ at more than $2\,\sigma$. In figure
\ref{fig:data1} and figure \ref{fig:data2}, we plot the theoretical
best fit model $f_{\rm NL}=48$ of our non-Gaussianity calculations
(red solid lines). These curves match the observed power spectra
very well, especially on large scales.

\subsection{Other Shapes}

Generally, the correction to the standard halo bias due to the
presence of primordial non-Gaussianity is $\Delta b_{\rm h}/b_{\rm
h}=\delta_{\rm c}\beta(k)$, where the function $\beta(k)$ is given
by \cite{liciaapjl}:
\begin{equation}
\beta(k)=\frac{1.3}{8\pi^2\sigma^2_{\rm M}\mathcal{M}_{\rm
R}(k)}\int{dk_1}k_1\mathcal{M}_{\rm
R}(k_1)\int^1_{-1}d\mu\mathcal{M}_{\rm
R}(\sqrt{\alpha})\frac{B_{\Phi}(k_1,\sqrt{\alpha},k)}{P_\Phi(k)}~,
\end{equation}
where $\mathcal{M}_{\rm R}=4\pi\rho_{\rm m}R^3/3$ is the halo mass
which is related to the scale $R$, $B_\Phi(k)$ denotes the
expression for the primordial bispectrum of the Bardeen potential
$\Phi$, $P_\Phi$ its power spectrum and $\alpha\equiv k_1^2 + k^2 +
2k_1 k \mu$.

In the local non-Gaussian case, the expression for the  correction
to the standard gaussian halo bias simplify [eq.
(\ref{eq:nghalobias})]. In this subsection, we explore other two
different types of primordial non-Gaussianity  given by  the
Equilateral and Enfolded templates.

The equilateral type of non-Gaussianity
\cite{Seery:2005wm,Chen:2006nt}, which can be generated from the
inflationary models with higher-derivative operators of the
inflaton, can be well described by the following template
\cite{Creminelli:2005hu}: $B_\Phi(k_1,k_2,k_3)=6f_{\rm NL}^{\rm
eq}F^{\rm eq}(k_1,k_2,k_3)$, where
\begin{equation}
F^{\rm
eq}(k_1,k_2,k_3)=(-P_\Phi(k_1)P_\Phi(k_2)+2cyc)-2[P_\Phi(k_1)P_\Phi(k_2)P_\Phi(k_3)]^{2/3}+(P_\Phi(k_1)^{1/3}P_\Phi(k_2)^{2/3}P_\Phi(k_3)+5cyc)~.
\end{equation}
Then we find that the correction $\beta(k)$ in the Equilateral
non-Gaussian case is almost scale-independent (see figure 1 of ref.
\cite{liciaapjl}). In principle there is a mass dependence (see
discussion in ref. \cite{Wagner2}), but the template and the actual
physical bispectrum yield different effective $f_{\rm NL}$ (denoted
by $\tilde{f}_{\rm NL}$) for the halo bias. Here, therefore, we
consider a phenomenological scale-independent scaling and neglect
the mass dependence. The interpretation of the resulting constraint
should be interpreted as a constraint on an effective  $f_{\rm NL}$,
its connection to the physical bispectrum depends on the detailed
model of inflation under consideration.

Using all datasets, we obtain the constraint on the Equilateral type
of primordial non-Gaussianity:
\begin{equation}
\tilde{f}^{\rm eq}_{\rm NL}=50\pm265~~(1\,\sigma)~,~~~~-419 <
\tilde{f}^{\rm eq}_{\rm NL} < 625~~(2\,\sigma)~,
\end{equation}
which is consistent with the WMAP7 results \cite{WMAP7}.

The template of another type of non-Gaussianity ``Enfolded'', which
arises in models with non-Bunch-Davies initial state
\cite{Chen:2006nt,Holman:2007na} or in effective field theories of
inflation with higher-derivative interactions \cite{BFMR1,BFMR2}
(see also ref.~\cite{galileon}), is given by ref.
\cite{Meerburg:2009ys}: $B_\Phi(k_1,k_2,k_3)=6f_{\rm NL}^{\rm
enf}F^{\rm enf}(k_1,k_2,k_3)$, where
\begin{equation}
F^{\rm
enf}(k_1,k_2,k_3)=(P_\Phi(k_1)P_\Phi(k_2)+2cyc)+3[P_\Phi(k_1)P_\Phi(k_2)P_\Phi(k_3)]^{2/3}-(P_\Phi(k_1)^{1/3}P_\Phi(k_2)^{2/3}P_\Phi(k_3)+5cyc)~.
\end{equation}
In figure 1 of ref. \cite{liciaapjl}, it can be seen that the
correction in this type is proportion to $k^{-1}$ on large scales.

No single field inflationary model has this scaling in the squeezed
limit: all inflationary single field models considered so far have
only the local or the template equilateral scaling in this limit
\cite{Wagner2}. This is nevertheless an interesting intermediate
case to consider. In fact non-single field models can yield all the
intermediate scalings in the squeezed limit, see e.g., ref.
\cite{Chen10}.  We therefore consider a phenomenological scaling
like $k^{-1}$ which could cover this Enfolded  template case and
neglect the mass dependence.  In this case, the whole data sets give
the constraint on the effective non-Gaussianity parameter for the
enfolded template:
\begin{equation}
\tilde{f}^{\rm enf}_{\rm NL}=183\pm95~~(1\,\sigma)~,~~~~-12 < \tilde{f}^{\rm enf}_{\rm NL} <
358~~(2\,\sigma)~.
\end{equation}
In the  WMAP7 paper \cite{WMAP7}, a constraint on the Orthogonal
type of non-Gaussianity \cite{Chen:2006nt} is given. The correction
to the halo bias of this  template is also proportional to $k^{-1}$
on large scales, but have the opposite sign and  a normalization
factor of $\tilde{f}_{\rm NL}^{\rm orth}=-2 \tilde{f}_{\rm NL}^{\rm
enf}$ with a weak mass dependence which we neglect here. Therefore,
based on the result of Enfolded template, in our simple analysis, we
obtain the constraint on the non-Gaussianity parameter for the
orthogonal template: $\tilde{f}^{\rm orth}_{\rm NL}=-92\pm 47$
($1\,\sigma$) and $-179 < \tilde{f}^{\rm orth}_{\rm NL} < 6$
($2\,\sigma$). The orthogonal type of non-Gaussianity in the
squeezed limit however  scales  like the  equilateral one and, as
shown in ref. \cite{Wagner2}  its effect on the halo bias is almost
degenerate with the equilateral non-Gaussianity. The enfolded type
of non-Gaussianity arising for example from modified initial state,
in the squeezed limit scales like the local template and as shown in
ref. \cite{Wagner2}  its effect on the halo bias is almost
degenerate with the local non-Gaussianity but with a different
normalization $f_{\rm NL}^{\rm mod.in.state}\simeq 1/8 f_{\rm
NL}^{\rm local}$. This re-scaling can be used to re-interpret the
local constraints into constraints for the modified initial state
non-Gaussianity.

\subsection{Cubic Correction $g_{\rm NL}$}

Finally, we consider the effect of the cubic correction on the halo
bias model. In this case, the Bardeen potential $\Phi$ can be
conveniently parameterized up to third order by:
\begin{equation}
\Phi=\phi+f_{\rm NL}(\phi^2-\langle\phi^2\rangle)+g_{\rm NL}\phi^3~,
\end{equation}
where $g_{\rm NL}$ is dimensionless, phenomenological parameter.
Ref. \cite{seljakgnl} explored the effect of a cubic correction
$g_{\rm NL}\phi^3$ on the mass function and halo bias model in
detail and gave the expression of the scale-dependent bias
correction when $f_{\rm NL}=0$ and $g_{\rm NL}\neq0$:
\begin{equation}
\Delta b(k,g_{\rm NL})=\frac{1}{4}\epsilon_{\rm k}g_{\rm
NL}\delta_{\rm c}(z)S^{(1)}_{\rm 3,M}\Delta b(k,f_{\rm NL}=1)~,
\end{equation}
where the normalized skewness of the density field $S^{(1)}_{\rm
3,M}=S_{\rm 3,M}/f_{\rm NL}$ and the parameter $\epsilon_{\rm
k}\simeq0.6$ obtained by the fitting in ref. \cite{seljakgnl}. Then
we obtain the constraint on $g_{\rm NL}$ from all the data sets:
\begin{equation}
g_{\rm NL}=(5.7\pm3.0)\times10^5~~(1\,\sigma)~,~~~~-1.2\times10^5 < g_{\rm NL} <
11.3\times10^5~~(2\,\sigma)~.
\end{equation}
The result is compatible with that obtained in refs.
\cite{seljakgnl,Fergusson:2010gn}.

\begin{table}
\caption{$1,\,2\,\sigma$ constraints on the primordial
non-Gaussianity from different data combinations. We report the mean
values and the Bayesian central credible interval, marginalized over
all other parameters.} \label{tab:I}
\begin{center}
\begin{tabular}{c|c|c}
\hline \hline

Datasets&\multicolumn{2}{c}{Non-Gaussianity}\\

\hline

WMAP7+BAO+SN&~~~~$1\,\sigma$~~~~~~~&~~~~$2\,\sigma$~~~~~~~\\

\hline

\multicolumn{3}{c}{Local Type $f_{\rm NL}$}\\

\hline
$+C_\ell^{\rm XX}$(NVSS)&$78\pm52$&$[-34,187]$\\
$+C_\ell^{\rm XX}+C_\ell^{\rm XT}$(NVSS)&$74\pm40$&$[-16,166]$\\

\hline

$+C_\ell^{\rm XX}$(QSO)&$62\pm26$&$[5,115]$\\
$+C_\ell^{\rm XX}+C_\ell^{\rm XT}$(QSO)&$59\pm21$&$[17,103]$\\

\hline

$+C_\ell^{\rm XX}$(LRG)&$165\pm105$&$[-81,351]$\\
$+C_\ell^{\rm XX}+C_\ell^{\rm XT}$(LRG)&$153\pm95$&$[-51,347]$\\

\hline

$+C_\ell^{\rm XX}$(ALL)&$68\pm22$&$[22,108]$\\
$+C_\ell^{\rm XX}+C_\ell^{\rm XY}+C_\ell^{\rm XT}$(ALL)&$48\pm20$&$[5,84]$\\

\hline\hline

\multicolumn{3}{c}{Equilateral Template $\tilde{f}_{\rm NL}$}\\

\hline
$+C_\ell^{\rm XX}+C_\ell^{\rm XY}+C_\ell^{\rm XT}$(ALL)&$50\pm265$&$[-419,625]$\\

\hline\hline

\multicolumn{3}{c}{Enfolded Template $\tilde{f}_{\rm NL}$}\\

\hline
$+C_\ell^{\rm XX}+C_\ell^{\rm XY}+C_\ell^{\rm XT}$(ALL)&$183\pm95$&$[-12,358]$\\

\hline\hline

\multicolumn{3}{c}{Cubic Correction $g_{\rm NL}\times10^{-5}$}\\

\hline
$+C_\ell^{\rm XX}+C_\ell^{\rm XY}+C_\ell^{\rm XT}$(ALL)&$5.7\pm3.0$&$[-1.2,11.3]$\\

\hline \hline
\end{tabular}
\end{center}
\end{table}

\begin{figure}
\begin{center}
\includegraphics[scale=0.4]{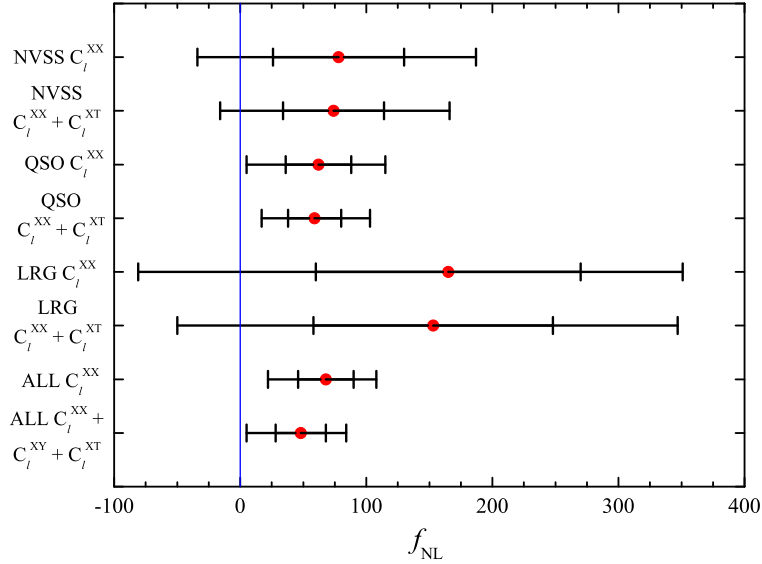}
\caption{Median values (red points) and $1\,\sigma$ and $2\,\sigma$
limits on the local type of primordial non-Gaussianity $f_{\rm NL}$
obtained from different data combinations. The vertical line denotes
the Gaussian case $f_{\rm NL}=0$.\label{fig:fnl}}
\end{center}
\end{figure}

\section{Summary \& Conclusions}
\label{summary}

Searches for primordial non-Gaussianity can provide key information
on the physical mechanism that generate the primordial perturbations
and can constrain specific inflationary models. A relatively new and
promising method to detect primordial non-Gaussianity  exploits the
fact that the clustering of dark matter halos (which host massive
galaxies and QSOs) is modified on large-scales in a scale-dependent
way. This effect is called non-Gaussian halo bias.  To access these
large scales, large-scale structure surveys covering large volumes,
high redshift and tracers of massive halos are most suited. Among
the currently available surveys, the NVSS and the SDSS QSO (in
particular the DR6 catalog) are the largest. Previous work had
detected excess large-scale power in their auto-correlation function
yielding a detection of a positive value of the non-Gaussianity
parameter $f_{\rm NL}$ at $> 2 \sigma$ level. Here we have revisited
the issue using the angular power spectrum rather than the
correlation function. While the two statistics should enclose the
same physical information they are affected by systematics effects
in different ways: this  complementary analysis offers thus a
consistency check and a test for possible systematics. We have also
used the recently updated MegaZ-LRG survey: a previous release of
the same survey  had also  been found  to have excess large-scale
power, but mostly at $\ell<10$. In an effort to be as conservative
as possible in our analysis, we excuded all multipoles at $\ell<10$:
these multipoles are affected by theoretical uncertainties and
systematics such as calibration, sample contamination etc. These
also probe the largest scales, thus these are the  multipoles where
most of the non-Gaussian  signal is  likely to be concentrated. We
combine the survey auto power spectra with the cross-power spectra
among the surveys and of each of the survey with the CMB temperature
map. In our analysis we marginalize over the LCDM cosmological
parameters with a  prior imposed by WMAP7 angular power spectra,
BAO, supernovae and Hubble constant measurement. We have
investigated the constraints on the parameter $f_{\rm NL}$,
characterizing primordial non-Gaussianity not only for the popular
local shape but also for the equilateral, enfolded and orthogonal
templates.  We have also considered constraints on a cubic
non-Gaussianity correction (parameterized by the $g_{\rm NL}$
parameter). We find no evidence for non-Gaussianity for any
``shape'' which halo bias effect on large scale scales less steeply
than $1/k^2$ and no evidence for non-zero $g_{\rm NL}$.  However for
the local type of non-Gaussianity we find a $\sim 2\sigma$ signal
for positive $f_{\rm NL}$ when all data-sets are combined. The
results are summarized in figure \ref{fig:fnl}: no single data set
drives the signal and the central $f_{\rm NL}$ values recovered are
fully in agreement with previous analyses.  Due to the more
conservative approach taken here (and thus larger error-bars), the
only  single survey that yield more than $2\sigma$ detection is the
SDSS DR7 QSO survey. The NVSS survey  central recovered $f_{\rm NL}$
value is also in agreement with both the QSO measurement and
previous analyses but the error-bar is larger. The combination of
the data sets yields: $f_{\rm NL}=48\pm 20$ and $5<f_{\rm NL}<84$ at
$1$ and $2\,\sigma$. Thus the tantalizing hint for a positive local
$f_{\rm NL}$ found from the auto-correlation function remains
despite the different analysis. We have investigated several
possible sources of systematic errors such as the sources number
density and the contamination by Galactic emissions for NVSS
sources; different choices for the CMB temperature fluctuations
templates; possible contamination of stars for the SDSS and LRG
samples. Our main results are stable when these systematic errors
are considered. Such a result would have profound implications for
inflationary mechanisms. The improved statistical power of
forthcoming Large-Scale Structure surveys is needed to fully resolve
the issue. Moreover, it will be also important to use the CMB data
which will be provided by Planck.

\section*{Acknowledgements}

We acknowledge the use of the Legacy Archive for Microwave
Background Data Analysis and the HEALPix package. Numerical analysis
has been performed at the University of Cambridge High Performance
Computing Service (http://www.hpc.cam.ac.uk/). This research has
been partially supported by the ASI contract No. I/016/07/0 COFIS,
the ASI/INAF agreement I/072/09/0 for the Planck LFI Activity of
Phase E2 and a PRIN MIUR. MV is supported by ASI/AAE, a PRIN MIUR, a
PRIN INAF 2009 and the ERC-StG ``cosmoIGM''. Support was given by
the Italian Space Agency through the ASI contracts Euclid-IC
(I/031/10/0). LV acknowledges support from grant FP7 ERC-IDEAS
Phys.LSS 240117 and MICINN grant AYA2008-03531. LV and SM thank the
Centro de Ciencias de Benasque Pedro Pascual where stimulating
discussions about this work took place. Funding for the SDSS and
SDSS-II has been provided by the Alfred P. Sloan Foundation, the
Participating Institutions, the National Science Foundation, the
U.S. Department of Energy, the National Aeronautics and Space
Administration, the Japanese Monbukagakusho, the Max Planck Society,
and the Higher Education Funding Council for England. The SDSS Web
Site is http://www.sdss.org/. The SDSS is managed by the
Astrophysical Research Consortium for the Participating
Institutions. The Participating Institutions are the American Museum
of Natural History, Astrophysical Institute Potsdam, University of
Basel, University of Cambridge, Case Western Reserve University,
University of Chicago, Drexel University, Fermilab, the Institute
for Advanced Study, the Japan Participation Group, Johns Hopkins
University, the Joint Institute for Nuclear Astrophysics, the Kavli
Institute for Particle Astrophysics and Cosmology, the Korean
Scientist Group, the Chinese Academy of Sciences (LAMOST), Los
Alamos National Laboratory, the Max-Planck-Institute for Astronomy
(MPIA), the Max- Planck-Institute for Astrophysics (MPA), New Mexico
State University, Ohio State University, University of Pittsburgh,
University of Portsmouth, Princeton University, the United States
Naval Observatory, and the University of Washington.


\begin{thebibliography}{nn}

%Introduction==============================

\bibitem{Komatsuwhitepaper}
E. Komatsu {\it et al.}, The Astronomy and Astrophysics Decadal
Survey, 2010, 158 (2009).

\bibitem{bmr04}
N.~Bartolo, E.~Komatsu, S.~Matarrese and A.~Riotto, Phys.\ Rept.\
{\bf 402}, 103 (2004).

\bibitem{Komatsu2010}
E. Komatsu, Class.\ Quant.\ Grav.\  {\bf 27}, 124010 (2010).

\bibitem{Salopekbond90}
D.~S.~Salopek and J.~R.~Bond, Phys.\ Rev.\ D {\bf 42}, 3936 (1990).

\bibitem{Ganguietal94}
A.~{Gangui}, F.~{Lucchin}, S.~{Matarrese} and S.~{Mollerach},
Astrophy.\ J.\ {\bf 430}, 447 (1994).

\bibitem{VWHK00}
L.~{Verde}, L.~{Wang}, A.~F. {Heavens} and M.~{Kamionkowski}, Mon.\
Not.\ Roy.\ Astron.\ Soc.\ {\bf 313}, 141 (2000).

\bibitem{KS01}
E.~{Komatsu} and D.~N. {Spergel}, Phys.\ Rev.\ D {\bf 63}, 063002
(2001).

\bibitem{BabichCreminelliZaldarriaga}
D.~{Babich}, P.~{Creminelli} and M.~{Zaldarriaga}, JCAP {\bf 0408},
009 (2004).

\bibitem[{Dalal {et~al.} (2008)}]{DDHS08}
N.~Dalal, O.~Dor{\'e}, D.~Huterer and A.~Shirokov, Phys.\ Rev.\ D
{\bf 77}, 123514 (2008).

\bibitem[{Matarrese \& Verde (2008)}]{MV08}
S.~Matarrese and L.~Verde, Astrophy.\ J.\ {\bf 677}, L77 (2008).

\bibitem{gp10}
T.~Giannantonio, C.~Porciani, Phys.\ Rev.\ D {\bf 81}, 063530
(2010).

\bibitem{melita}
C.~Carbone, L.~Verde and S.~Matarrese, Astrophy.\ J.\ {\bf 684}, L1
(2008).

\bibitem{viel09}
M.~Viel, E.~Branchini, K.~Dolag, M.~Grossi, S.~Matarrese and
L.~Moscardini, Mon.\ Not.\ Roy.\ Astron.\ Soc.\ {\bf 393}, 774
(2009).

\bibitem{Condon:1998iy}
J.~J.~Condon, W.~D.~Cotton, E.~W.~Greisen, Q.~F.~Yin, R.~A.~Perley,
G.~B.~Taylor and J.~J.~Broderick, Astron.\ J.\ {\bf 115}, 1693
(1998).

\bibitem{richards09}
G.~T.~Richards {\it et al.}, Astrophys.\ J.\ Suppl.\ {\bf 180}, 67
(2009).

\bibitem{hoetal08}
S.~Ho, C.~Hirata, N.~Padmanabhan, U.~Seljak and N.~Bahcall, Phys.\
Rev.\ D {\bf 78}, 043519 (2008).

\bibitem{ISW}
R.~K.~Sachs and A.~M.~Wolfe, Astrophys.\ J.\ {\bf 147}, 73 (1967).

\bibitem{Slosar}
A.~Slosar, C.~Hirata, U.~Seljak, S.~Ho and N.~Padmanabhan, JCAP {\bf
0808}, 031 (2008).

\bibitem{afshorditolley}
N.~Afshordi and A.~J.~Tolley, Phys.\ Rev.\ D {\bf 78}, 123507
(2008).

\bibitem{xiaACF}
J.~Q.~Xia, M.~Viel, C.~Baccigalupi, G.~De Zotti, S.~Matarrese and
L.~Verde, Astrophy.\ J.\ {\bf 717}, L17 (2010).

\bibitem{xiaCCF}
J.~Q.~Xia, A.~Bonaldi, C.~Baccigalupi, G.~De Zotti, S.~Matarrese,
L.~Verde and M.~Viel, JCAP {\bf 1008}, 013 (2010).

\bibitem{LRG7}
S.~A.~Thomas, F.~B.~Abdalla and O.~Lahav, arXiv:1011.2448.

\bibitem{Gnlpaper}
N.~Bartolo, S.~Matarrese and A.~Riotto, Phys.\ Rev.\ D {\bf 69},
043503 (2004).

\bibitem{loverdeetal08}
M.~LoVerde, A.~Miller, S.~Shandera and L.~Verde, JCAP {\bf 0804},
014 (2008).

%====================================================================



%\bibitem{Overzier}
%R.~A.~Overzier, H.~J.~A.~R{\"o}ttgering, R.~B.~Rengelink and
%R.~J.~Wilman, Astron.\ Astrophys.\ {\bf 405}, 53 (2003).








%Theoretical Framework=====================

%\bibitem{afshordietal04}
%N.~Afshordi, Y.~S.~Loh, and M.~A.~Strauss, Phys.\ Rev.\ D {\bf 69},
%083524 (2004).

%\bibitem{peiris00}
%H.~V.~Peiris and D.~N.~Spergel, Astrophys.\ J.\ {\bf 540}, 605
%(2000).

%\bibitem{cooray02}
%A.~Cooray, Phys.\ Rev.\ D {\bf 65}, 103510 (2002).


\bibitem{shethtormen}
R.~K.~Sheth and G.~Tormen, Mon.\ Not.\ Roy.\ Astron.\ Soc.\ {\bf
308}, 119 (1999).

\bibitem{mvj}
S.~Matarrese, L.~Verde and R.~Jimenez, Astrophys.\ J.\ {\bf 541}, 10
(2000).

\bibitem{VJKM01}
L.~{Verde}, R.~{Jimenez}, M.~{Kamionkowski} and S.~{Matarrese},
Mon.\ Not.\ Roy.\ Astron.\ Soc.\ {\bf 325}, 412 (2001).

\bibitem{MR10}
M.~Maggiore and A.~Riotto, Astrophys.\ J.\ {\bf 717} (2010) 526.

\bibitem{D'Amico:2010ta}
G.~D'Amico, M.~Musso, J.~Norena and A.~Paranjape, JCAP {\bf 1102},
001 (2011).


\bibitem{grossi09}
M.~Grossi {\it et al.}, Mon.\ Not.\ Roy.\ Astron.\ Soc.\ {\bf 398},
321 (2009).

\bibitem{wagner1}
C.~Wagner, L.~Verde and L.~Boubekeur, JCAP {\bf 1010}, 022 (2010).

\bibitem{desjacque09}
V.~Desjacques, U.~Seljak and I.~T.~Iliev, Mon.\ Not.\ Roy.\ Astron.\
Soc.\ {\bf 396}, 85 (2009).

\bibitem{Smith:2009jr}
K.~M.~Smith, L.~Senatore and M.~Zaldarriaga, JCAP {\bf 0909}, 006
(2009).

\bibitem{Wagner2}
C.~Wagner and L.~Verde, arXiv:1102.3229.

%\bibitem{Creminelli}
%N.~Creminelli {\it et al.}, in preparation.

\bibitem{Chen10}
X.~Chen and Y.~Wang, Phys.\ Rev.\ D {\bf 81}, 063511 (2010).

\bibitem{Mo96}
H.~J.~Mo and S.~D.~M.~White, Mon.\ Not.\ Roy.\ Astron.\ Soc.\ {\bf
282}, 347 (1996).

\bibitem{Mo97}
H.~J.~Mo, Y.~P.~Jing and S.~D.~M.~White, Mon.\ Not.\ Roy.\ Astron.\
Soc.\ {\bf 284}, 189 (1997).

\bibitem{Scoccimarro01}
R.~Scoccimarro, R.~Sheth, L.~Hui and B.~Jain, Astrophys.\ J.\ {\bf
546}, 20 (2001).

%\bibitem{WallJenkins}
%J.~V.~Wall and C.~R.~Jenkins, Practical Statistics for Astronomers,
%Cambridge University Press (2003).

\bibitem{BlakeWall}
C.~Blake and J.~Wall, Mon.\ Not.\ Roy.\ Astron.\ Soc.\ {\bf 337},
993 (2002).

\bibitem{BlakeWall2}
C.~Blake and J.~Wall, Mon.\ Not.\ Roy.\ Astron.\ Soc.\ {\bf 329},
L37 (2002).


\bibitem[Hernandez-Monteagudo(2010)]{CHM10}
C.~Hernandez-Monteagudo, Astron.\ Astrophys.\ {\bf 520}, 101 (2010).

%NVSS Radio Sources========================

\bibitem{Brookes}
M.~H.~Brookes, P.~N.~Best, J.~A.~Peacock, H.~J.~A.~R\"otgering and
J.~S.~Dunlop, Mon.\ Not.\ Roy.\ Astron.\ Soc.\ {\bf 385}, 1297
(2008).

\bibitem{DeZotti10}
G.~de Zotti, M.~Massardi, M.~Negrello and J.~Wall, Astron.\
Astrophys.\ Rev.\ {\bf 18}, 1 (2010).

\bibitem{xiaISW}
J.~Q.~Xia, M.~Viel, C.~Baccigalupi and S.~Matarrese, JCAP {\bf
0909}, 003 (2009).

\bibitem{myersetal06}
A.~D.~Myers {\it et al.}, Astrophys.\ J.\ {\bf 638}, 622 (2006).

\bibitem{gianna06}
T.~Giannantonio {\it et al.}, Phys.\ Rev.\ D {\bf 74}, 063520
(2006).

\bibitem{gianna08}
T.~Giannantonio {\it et al.}, Phys.\ Rev.\ D {\bf 77}, 123520
(2008).

\bibitem{LRGredshift}
A.~Collister {\it et al.}, Mon.\ Not.\ Roy.\ Astron.\ Soc.\ {\bf
375}, 68 (2007).

%================================================================

%\bibitem{peebles}
%P.~J.~E.~Peebles, Astrophys.\ J.\ {\bf 185}, 413 (1973).

\bibitem{Hivon}
E.~Hivon {\it et al.}, Astrophys.\ J.\ {\bf 567}, 2 (2002).

\bibitem{peeblesGll}
M.~G.~Hauser and P.~J.~E.~Peebles, Astrophys.\ J.\ {\bf 185}, 757
(1973).

\bibitem{healpix}
K.~M.~G\'orski {\it et al.}, Astrophys.\ J.\ {\bf 622}, 759 (2005).

\bibitem{Yoo1}
J.~Yoo, A.L.~Fitzpatrick and M.~Zaldarriaga, Phys.\ Rev.\ D {\bf
80}, 083514 (2009).

\bibitem{Yoo2}
J.~Yoo, Phys.\ Rev.\ D {\bf 82}, 083508 (2010)

\bibitem{Hinshaw}
G.~Hinshaw {\it et al.}, Astrophys.\ J.\ Suppl.\ {\bf 148}, 135
(2003).

\bibitem{planck}
Planck Collaboration, arXiv:1101.2022.

\bibitem{ilc7}
B.~Gold {\it et al.}, Astrophys.\ J.\ Suppl.\ {\bf 192}, 15 (2011).

\bibitem{Jarosik11}
N.~Jarosik {\it et al.}, Astrophys.\ J.\ Suppl.\ {\bf 192}, 14
(2011).

\bibitem{nobs}
M.~Limon {\it et al.}, ``Wilkinson Microwave Anisotropy Probe
(WMAP): Seven Year Explanatory Supplement''.

%================================================================

\bibitem{camb}
A.~Lewis, A.~Challinor and A.~Lasenby, Astrophys.\ J.\ {\bf 538},
473 (2000).

\bibitem{Halofit}
R.~E.~Smith {\it et al.}, Mon.\ Not.\ Roy.\ Astron.\ Soc.\ {\bf
341}, 1311 (2003).

%\bibitem{Negrello}
%M.~Negrello, M.~Magliocchetti and G.~De~Zotti, Mon.\ Not.\ Roy.\
%Astron.\ Soc.\ {\bf 368}, 935 (2006).

%\bibitem{ws09}
%D.~Wands and A.~Slosar, Phys.\ Rev. D, {\bf 79}, 123507 (2009).


%\bibitem{Delabrouille09}
%J.~Delabrouille, J.~F.~Cardoso, M.~Le~Jeune, M.~Betoule, G.~Fay and
%F.~Guilloux, Astron.\ Astrophys.\ {\bf 493}, 835 (2009).

%\bibitem{covariance}
%R.~Scranton {\it et al.}, Astrophys.\ J.\ {\bf 579}, 48 (2002).


%\bibitem[{Raccanelli {et~al.} (2008)}]{Raccanelli}
%A.~Raccanelli {\it et al.}, Mon.\ Not.\ Roy.\ Astron.\ Soc.\ {\bf
%386}, 2161 (2008).

%\bibitem{Massardi10}
%M.~Massardi, A.~Bonaldi, M.~Negrello, S.~Ricciardi, A.~Raccanelli
%and G.~De~Zotti, Mon.\ Not.\ Roy.\ Astron.\ Soc.\ {\bf 404}, 532
%(2010).

%SDSS DR6 Quasars==========================

%\bibitem{richardsetal04}
%G.~T.~Richards {\it et al.}, Astrophys.\ J.\ Suppl.\ {\bf 155}, 257
%(2004).


%Anna ============
%\bibitem{synch}
% C. G. Haslam, C. J.Salter,  H. Stoffel,  W. E. Wilson, 1982, A\&AS 47, 1

%\bibitem{dust}
% D.~J. Schlegel, D.~P.Finkbeiner, M.Davis, 1998, ApJ v.500, p.525

%\bibitem{ff}
%C. Dickinson, R.~D. Davies, and R.~J. Davis, 2003,  Mon.\ Not.\ Roy.\ Astron.\ Soc. 341, 369


%Method & Datasets=========================

\bibitem{cosmomc}
A.~Lewis and S.~Bridle, Phys.\ Rev.\ D {\bf 66}, 103511 (2002); URL:
http://cosmologist.info/cosmomc/.

\bibitem{liciachi2}
L.~Verde {\it et al.}, Astrophys.\ J.\ Suppl.\ {\bf 148}, 195
(2003).

\bibitem{WMAP7}
E.~Komatsu {\it et al.}, Astrophys.\ J.\ Suppl.\  {\bf 192}, 18
(2011).

\bibitem{BAO}
W.~J.~Percival {\it et al.}, Mon.\ Not.\ Roy.\ Astron.\ Soc.\ {\bf
401}, 2148 (2010).

\bibitem{Eisenstein:2005su}
D.~J.~Eisenstein {\it et al.} Astrophys.\ J.\  {\bf 633}, 560
(2005).

\bibitem{Eisenstein:1997ik}
D.~J.~Eisenstein and W.~Hu, Astrophys.\ J.\  {\bf 496}, 605 (1998).

\bibitem{Amanullah:2010vv}
R.~Amanullah {\it et al.}, Astrophys.\ J.\  {\bf 716}, 712 (2010).

\bibitem{Kowalski:2008ez}
M.~Kowalski {\it et al.}, Astrophys.\ J.\  {\bf 686}, 749 (2008).

\bibitem{Amanullah:2007yv}
R.~Amanullah {\it et al.}, Astron.\ Astrophys.\ {\bf 486}, 375
(2008).

\bibitem{Holtzman08}
J.~A.~Holtzman {\it et al.}, Astron.\ J.\  {\bf 136}, 2306 (2008).

\bibitem{Hicken09}
M.~Hicken {\it et al.}, Astrophys.\ J.\  {\bf 700}, 331 (2009).

\bibitem{SNMethod1}
M.~Goliath, R.~Amanullah, P.~Astier, A.~Goobar and R.~Pain, Astron.\
Astrophys.\ {\bf 380}, 1 (2001).

\bibitem{SNMethod2}
E.~Di~Pietro and J.~F.~Claeskens, Mon.\ Not.\ Roy.\ Astron.\ Soc.\
{\bf 341}, 1299 (2003).

\bibitem{HST}
A.~G.~Riess {\it et al.}, Astrophys.\ J.\ {\bf 699}, 539 (2009).

\bibitem{Croom}
S.~M.~Croom {\it et al.}, Mon.\ Not.\ Roy.\ Astron.\ Soc.\ {\bf
356}, 415 (2005).

%\bibitem{Gold}
%B.~Gold {\it et al.}, arXiv:1001.4555.

%\bibitem{Stivoli}
%F.~Stivoli, J.~Grain, S.~M.~Leach, M.~Tristram, C.~Baccigalupi and
%R.~Stompor, arXiv:1004.4756.

%\bibitem{Bonaldi}
%A.~Bonaldi, S.~Ricciardi, S.~Leach, F.~Stivoli, C.~Baccigalupi and
%G.~de~Zotti, Mon.\ Not.\ Roy.\ Astron.\ Soc.\ {\bf 382}, 1791
%(2007).

%\bibitem[Hickox et al.(2009)]{Hickox} Hickox, R.~C., et al.\
%2009, Astrophys.\ J., 696, 891

%Numerical Results=========================

\bibitem{blakeNVSS}
C.~Black, P.~G.~Ferreira and J.~Borrill, Mon.\ Not.\ Roy.\ Astron.\
Soc.\ {\bf 351}, 923 (2004).

\bibitem{blakeLRG}
C.~Blake, A.~Collister, S.~Bridle and O.~Lahav, Mon.\ Not.\ Roy.\
Astron.\ Soc.\ {\bf 374}, 1527 (2007).

\bibitem{LRGPRL}
S~.A.~Thomas, F.~B.~Abdalla and O.~Lahav, arXiv:1012.2272.

\bibitem{PadmanabhanLRG}
N.~Padmanabhan {\it et al.}, Mon.\ Not.\ Roy.\ Astron.\ Soc.\ {\bf
378}, 852 (2007).

\bibitem{Fantaye}
Y.~Fantaye {\it et al.}, in preparation.

\bibitem{Yadav:2007yy}
A.~P.~S.~Yadav and B.~D.~Wandelt, Phys.\ Rev.\ Lett.\ {\bf 100},
181301 (2008).

\bibitem{Pietrobon:2008ve}
D.~Pietrobon, P.~Cabella, A.~Balbi, G.~de Gasperis and N.~Vittorio,
Mon.\ Not.\ Roy.\ Astron.\ Soc.\ {\bf 396}, 1682 (2009).


\bibitem{Curto:2009pv}
A.~Curto, E.~Martinez-Gonzalez and R.~B.~Barreiro, Astrophys.\ J.\
{\bf 706}, 399 (2009).

\bibitem{Smidt:2009ir}
J.~Smidt, A.~Amblard, P.~Serra and A.~Cooray, Phys.\ Rev.\ D {\bf
80}, 123005 (2009).

\bibitem{JV}
R.~Jimenez and L.~Verde, Phys.\ Rev.\ D {\bf 80}, 127302 (2009).

\bibitem{Rud}
O.~Rudjord, F.~K.~Hansen, X.~Lan, M.~Liguori, D.~Marinucci and
S.~Matarrese, Astrophys.\ J.\ {\bf 701}, 369 (2009).

\bibitem{liciaapjl}
L.~Verde and S.~Matarrese, Astrophys.\ J.\ {\bf 706}, L91 (2009).

\bibitem{Seery:2005wm}
D.~Seery and J.~E.~Lidsey, JCAP {\bf 0506}, 003 (2005).

\bibitem{Chen:2006nt}
X.~Chen, M.~X.~Huang, S.~Kachru and G.~Shiu, JCAP {\bf 0701}, 002
(2007).

\bibitem{Creminelli:2005hu}
P.~Creminelli, A.~Nicolis, L.~Senatore, M.~Tegmark and
M.~Zaldarriaga, JCAP {\bf 0605}, 004 (2006).

\bibitem{Holman:2007na}
R.~Holman and A.~J.~Tolley, JCAP {\bf 0805}, 001 (2008).

\bibitem{BFMR1}
N.~Bartolo, M.~Fasiello, S.~Matarrese and A.~Riotto, JCAP {\bf
1008}, 008 (2010).

\bibitem{BFMR2}
N.~Bartolo, M.~Fasiello, S.~Matarrese and A.~Riotto, JCAP {\bf
1012}, 026 (2010).

\bibitem{galileon}
P.~Creminelli, G.~D'Amico, M.~Musso, J.~Norena and E.~Trincherini,
JCAP {\bf 1102}, 006 (2011).

\bibitem{Meerburg:2009ys}
P.~D.~Meerburg, J.~P.~van der Schaar and P.~S.~Corasaniti, JCAP {\bf
0905}, 018 (2009).


\bibitem{seljakgnl}
V.~Desjacques and U.~Seljak, Phys.\ Rev.\ D {\bf 81}, 023006 (2010).

\bibitem{Fergusson:2010gn}
J.~R.~Fergusson, D.~M.~Regan and E.~P.~S.~Shellard, arXiv:1012.6039.




%Discussion & Conclusions==================


\end{thebibliography}
\end{document}